\documentclass[11pt]{article}
\usepackage{amssymb,cite,graphics}
\newcommand\blank[1]{}
\newcommand\ttl[1]{`#1', }
\newcommand\toline[1]{--#1}

\newcommand{\fract}[2]{{\textstyle\frac{#1}{#2}}}
\newcommand{\fr}[2]{{\textstyle\frac{#1}{#2}}}
\newcommand{\ri}{\right}
\newcommand{\ep}{\varepsilon}
\newcommand{\lf}{\left}
\newcommand{\te}{\theta}

\newcommand\ZZ{{\mathbb Z}}
\newcommand\NN{{\mathbb N}}
\newcommand\Zt{{\mathbb Z}_2}

\newcommand\ZN{{\mathbb Z}_N}

\newcommand{\CM}{{\cal M}}

\newcommand{\CK}{{\cal K}}
\newcommand\eq{\begin{equation}}
\newcommand\en{\end{equation}}
\newcommand\bea{\begin{eqnarray}}
\newcommand\eea{\end{eqnarray}}
\newcommand\nn{\nonumber}

\newcommand{\wt}{\tilde}
\newcommand{\FT}{{\cal F}}
\newcommand{\sss}[1]{\scriptscriptstyle{\rm #1}}
\newcommand\MR{M\!R}
\newcommand{\resection}[1]{\setcounter{equation}{0}\section{#1}}

\begin{document}
\begin{titlepage}
\vskip 0.5cm
\begin{flushright}
DTP-99/89 \\
T99/148 \\
ITFA 99/43\\
{\tt hep-th/0001185} \\
\end{flushright}
\vskip 1.5cm
\begin{center}
{\Large{\bf
New families of
flows between two-dimensional
conformal field theories
}}
\end{center}
\vskip 0.8cm
\centerline{Patrick Dorey%
\footnote{{\tt p.e.dorey@durham.ac.uk}},
Clare Dunning%
\footnote{{\tt t.c.dunning@durham.ac.uk}}
and Roberto Tateo%
\footnote{{\tt tateo@wins.uva.nl}}}
\vskip 0.9cm
\centerline{${}^1$\sl\small SPhT Saclay,
91191 Gif-sur-Yvette cedex, France\,}
\vskip 0.2cm
\centerline{${}^{1,2}$\sl\small Dept.~of Mathematical Sciences,
University of Durham, Durham DH1 3LE, UK\,}
\vskip 0.2cm
\centerline{${}^3$\sl\small Universiteit 
van Amsterdam, Inst.~voor Theoretische
Fysica, 1018 XE Amsterdam, NL\,}
\vskip 1.25cm
\begin{abstract}
\noindent

We present evidence for the existence of infinitely-many new families
of renormalisation group
flows between the nonunitary minimal models of conformal field theory.
These are associated with perturbations by the $\phi_{21}$ and $\phi_{15}$
operators, and generalise a family of flows discovered by Martins.
In all of the new flows, the finite-volume effective central charge is a
non-monotonic function of the system size. The evolution of this effective
central charge is studied by means of a nonlinear integral equation, a
massless variant of an equation recently found to describe certain
massive perturbations of these same models. We also observe that a
similar non-monotonicity arises in the more familiar 
$\phi_{13}$ perturbations, when the flows induced are between
nonunitary minimal models.

\end{abstract}
\end{titlepage}
\setcounter{footnote}{0}
\def\thefootnote{\fnsymbol{footnote}}
%
\resection{Introduction}
\label{intr}
Integrable quantum field theories 
which interpolate between different conformal
field theories have been recognised as an intriguing feature of
two-dimensional models ever since the initial paper of
A.B.~Zamolodchikov~\cite{Za}. The first examples 
arose as
$\phi_{13}$ perturbations of unitary minimal models $\CM_{p,p+1}$.
The existence of higher-spin integrals of motion\cite{Zb}
indicates that such perturbations should be integrable;
in~\cite{Za} (see also \cite{LC}) perturbative arguments
were used to show that for one sign of the coupling
constant the resulting flow is to the neighbouring
unitary minimal model,
namely $\CM_{p-1,p}\,$, at least for $p\gg 1$. Further support
for this picture, this time valid at all values of $p$, was
provided when ideas from the thermodynamic Bethe ansatz (TBA) enabled
exact integral equations describing the 
flows to be proposed~\cite{Zc}. 

Unitarity of a conformal field theory
is not a necessary requirement for the existence of
integrable perturbations, and in~\cite{La,Aa} the arguments of \cite{Za}
were extended to the
$\phi_{13}$ perturbations of more general minimal models $\CM_{p,q}$.
The results indicated that there should exist integrable flows between
the models $\CM_{p,q}$ and $\CM_{2p-q,p}$, at least in the region $p\gg
q{-}p$ where calculations perturbative in $(q{-}p)/p$ can be trusted.
(It remains an open problem
to give a TBA treatment of the nonunitary flows,
which would not suffer from this caveat.)
The unitary and nonunitary models can be put on a common footing by
defining a parameter $\zeta=p/(q{-}p)$\footnote{$\zeta$ was denoted $m$ 
in \cite{La}; the symbol $p$ is also often
used, but we are reserving this for the integer-valued first
index of the minimal models $\CM_{p,q}$}.
The value of $\zeta$ suffices to identify 
$p$ and $q$ uniquely, since they must be coprime.
In all cases the predicted flow is
from the model specified by $\zeta$ to the model specified by $\zeta{-}1$.

Many further massless flows have since been discovered and studied,
often by means of the TBA technique: the 
papers~\cite{Zab,FZa,KMspect,Maa,Ra1,Ma,RTV1,FOZ,FI,FQR,DTT,RT1,Rac} 
are a sample of this work.
Our main interest here will be the flows found by
Martins\cite{Ma} and further studied
by Ravanini et al.\cite{Rac}.  These are perturbations of the minimal models
$\CM_{p,2p-1}$ and $\CM_{p,2p+1}$ by the operators $\phi_{21}$ and
$\phi_{15}$ respectively, and the predicted trajectories are:
\bea
\phi_{21}~:~~\CM_{p,2p-1}&\to &\CM_{p-1,2p-1}~;
\label{m21flow}\\
\phi_{15}~:~~\CM_{p,2p+1}&\to &\CM_{p,2p-1}~.
\label{m15flow}
\eea
These flows can be chained together to form a single sequence, along
which the perturbing operator alternates between
$\phi_{21}$ and $\phi_{15}$.

In this paper we will extend this picture by proposing 
further sequences of flows for which the perturbing operator alternates in
the same way.
As in studies based on the TBA technique, our main tool will be a
conjecture for an
exact equation expressing the finite-volume ground state
energy of each model as a function
of the system size $R$. 
In fact, it will be most convenient to work with 
an associated scaling function known as the `effective central charge',
$c_{\rm eff}(r)$. This is related to the ground state energy as
\eq
E_0(M,R)=E_{\rm bulk}(M,R)-\frac{\pi}{6R}c_{\rm eff}(r)~,\quad
r=M\!R~.
\label{Egdef}
\en
In this equation, $E_0$ is the ground state energy, $E_{\rm bulk}$ the 
irregular bulk part, and 
$M$ some mass scale, set for example by an
infinite-volume one-particle state.
The effective central charge
depends on the system size only through the dimensionless
combination $M\!R$; for a model which happens to be scale invariant
(conformal), it is constant and
equal to $c{-}24d$, where $c$ is the model's central charge and
$d$ its lowest conformal dimension.
The previously-studied sequence (\ref{m21flow}), (\ref{m15flow}) is 
picked out by the monotonicity of the function $c_{\rm eff}$ as a
function of $r$. In all other cases, somewhat to our surprise, we found
that the effective central charge undergoes a number of oscillations as it
interpolates between its short and long distance limits.

Our approach differs from the TBA method, and is much closer to 
that of the papers~\cite{KBP,DDV,Zd,FMQR,FRT3,WBN,DT1}, 
in that a single nonlinear
integral equation (NLIE) is proposed to describe
infinitely-many different perturbed conformal field theories, each 
being picked out by an appropriate choice of certain parameters.
For massless flows, the one previous example of such an equation was
found by Al.~Zamolodchikov\cite{Zd}, and his paper formed
a large part of the motivation for our work.

The particular nonlinear integral equation 
that we will be using is obtained in \S2; then
in \S3 we discuss the nature of the flows that it predicts, and in \S4
we take a more detailed look at various asymptotics. This
allows us to back up our conjectures with a 
comparison with UV conformal perturbation theory.
In \S5 we
comment on some features of the $\phi_{13}$ flows, and in the concluding
\S6 we indicate some open problems that remain for future work.

\resection{The nonlinear integral equation}
\label{nlie}
The work of \cite{Zd} built in part on the nonlinear integral equation
of \cite{DDV}\footnote{the scale-invariant form of this equation
had arisen previously in the context of integrable lattice
models~\cite{KBP}}, which
encodes the finite-volume
ground state energy of the 
sine-Gordon model for a continuous range of the coupling $\beta$,
and at general `twist' $\alpha$. 
Setting
$\zeta=\beta^2/(8\pi{-}\beta^2)$, it can be shown that
for $\zeta=p/(q{-}p)$
and $\alpha=1/p$, the 
ground-state energy of the
$\phi_{13}$ perturbation of a general minimal model $\CM_{p,q}$ is also
matched~\cite{SmsG,lecl,ZamPIII,FMQR,FRT3}.
(Note that this definition of $\zeta$ is thus in line with the one
given in the introduction.) Since the sine-Gordon
model is massive, the minimal model perturbations reproduced in this way
are also massive. In contrast, the modification to the
equation found in \cite{Zd} describes a
massless perturbation of the sine-Gordon model, flowing from the model at
$\zeta$ to the model at $\zeta{-}1$,
and thus matching
the pattern of massless $\phi_{13}$ flows discussed in the
introduction. While only the $\alpha=0$ case was treated
explicitly in \cite{Zd}, we will confirm below that with a suitable
nonzero value of $\alpha$ the
massless $\phi_{13}$ perturbations of minimal models are also 
obtained, with results that agree with the perturbative picture. 
In the nonunitary cases this yields some genuinely new
information, since, as already mentioned,
TBA systems giving the exact evolution of the
ground-state energy for these models are not known.

However this is not the main purpose of our paper: rather, we would like to
perform a similar trick for the $\phi_{21}$ and $\phi_{15}$ perturbations
of minimal models. Recently, a nonlinear integral equation 
describing finite size effects in the $a_2^{(2)}$ model was
conjectured \cite{DT1}\footnote{as for
the sine-Gordon model, a scale invariant version of this equation had
previously been found in the context of integrable lattice models~\cite{WBN}},
and checked to describe the
massive
$\phi_{12}$ , $\phi_{21}$ and $\phi_{15}$ perturbations of minimal models
on the imposition of a suitably-chosen twist. (The connection between the
$a^{(2)}_2$ model and these perturbations can be understood through
quantum group reduction~\cite{Sm,MJM,GT}.)
Our aim is to modify this equation so as to find out about
the massless, interpolating, flows that may also be induced by these
perturbations.

The nonlinear integral equation found in \cite{DT1} is 
\eq
f(\te)=i \pi \alpha  -i r\sinh\theta
+\!\int_{{\cal
C}_1}\!\!\varphi(\te{-}\te')\ln(1{+}e^{f(\te')})\,d\te'
-\!\int_{{\cal
C}_2}\!\!\varphi(\te{-}\te')\ln(1{+}e^{-f(\te')})\,d\te'\,
\label{mnlie}
\en
with the effective central charge given in terms of its solution as
\eq
c_{\rm eff}(r)= {3 i r \over \pi^2} \lf(\int_{{\cal
C}_1} \sinh\theta\, \ln(1{+}e^{f(\te)})\,d\te-
\int_{{\cal C}_2}\sinh\theta\,\ln(1{+}e^{-f(\te)})\,d\te \ri)\,.
\label{mnliec}
\en
The contours ${\cal C}_1$
and ${\cal C}_2$ run from $-\infty$ to $+\infty$, 
just below and just above the real $\te$-axis,
and $r$ is equal to $\MR$,
$R$ being the size of the system and $M$ the mass scale, set by the
fundamental kink.
The kernel
\eq
\varphi(\te)=-\int_{-\infty}^{\infty}
\frac{e^{i k\te} \sinh(\fr{\pi}{3}k) \cosh( \fr{\pi}{6} k (1{-}2 \xi))} 
{\cosh( \fr{\pi}{2} k) \sinh( k \fr{\pi}{3} \xi )}%
\frac{d k}{2\pi}
\label{krnl}
\en
is equal to $i/2\pi$ times the logarithmic derivative of the scalar
factor in the S-matrix of the $a^{(2)}_2$
(Tzitz\'eica-Izergin-Korepin-Bullough-Dodd-Zhiber-Mikhailov-Shabat
\dots )
model. The parameter $\xi$, related to the $a^{(2)}_2$ coupling $\gamma$
as $\xi=\gamma/(2\pi-\gamma)$, is the analogue of the sine-Gordon
parameter $\zeta$ mentioned earlier, while $\alpha$ corresponds
to the
twist. To obtain massive
$\phi_{12}$, $\phi_{21}$ and $\phi_{15}$ perturbations of a minimal model
$\CM_{p,q}$, the values of $\xi$ and $\alpha$
must be chosen as follows~\cite{DT1}:
\bea
\phi_{12}~~&:&~~ \xi=\frac{1}{\fract{2q}{p}-1}~~\,,~~~\alpha=2/p
{}~~\,,~~~p<2q~;
\label{dt12p}\\
\phi_{21}~~&:&~~ \xi=\frac{1}{\fract{2p}{q}-1}~~\,,~~~\alpha=2/q
{}~~\,,~~~p>q/2~;
\label{dt21p}\\
\phi_{15}~~&:&~~ \xi=\frac{1}{\fract{q}{2p}-1}~~\,,~~~\alpha=1/p
{}~~\,,~~~p<q/2~;
\label{dt15p}
\eea
In each case, the inequality delimits the region of the $(p,q)$ plane
within which $\xi$ is positive, and
the corresponding perturbation is relevant.
(The first inequality is not strictly necessary, since we will be
adopting the convention that $p<q$ in the specification of
$\CM_{p,q}$.)
The UV effective central charge,
$c_{\rm eff}(0)=1-\frac{3\xi}{\xi+1}\alpha^2$, is always equal to $1-6/pq$, 
but the subsequent terms in the expansions
differ, as expected given the different perturbations being described.
 
The modification to the massive
equation of \cite{DDV} found in \cite{Zd} made use
of elements of the corresponding massless scattering
theory, proposed in \cite{FSZb}.
In our case we do not have a description of the massless scattering in terms 
of an S-matrix but, proceeding by analogy with the results of \cite{Zd},
we will substitute the single equation (\ref{mnlie})
with two equations describing hypothetical left and right movers.
{}From the formulae (\ref{dt21p}) and (\ref{dt15p}), we notice that
the massive versions of the
perturbations
(\ref{m21flow}) and (\ref{m15flow}) have $\xi=2p-1$ and
$\xi=2p$ respectively.
We therefore seek a
massless modification of the NLIE 
(\ref{mnlie}) which will interpolate in general 
between an ultraviolet theory with parameter $\xi$ and an infrared
theory with parameter
$\xi-1$.
To this end, we
introduce two analytic functions
$f_R(\te)$ and $f_L(\te)$, couple them together via
\bea
f_{R}(\te)&=& - i \frac{r}{2}e^{\te}  + i   \pi \alpha' 
\nn \\
&&+ \int_ {{\cal C}_1}\!\phi(\te-\te')
\ln(1+e^{f_{R}(\te')}) ~ d\te' 
- \int_ {{\cal C}_2}\! \phi(\te-\te')
\ln(1+e^{-f_{R}(\te')} ) ~ d\te'  
\nn \\
&&+  \int_ {{\cal C}_1}\!\chi(\te-\te')
\ln(1+e^{-f_{L}(\te')}) ~ d\te' 
- \int_ {{\cal C}_2}\! \chi(\te-\te')
\ln(1+e^{f_{L}(\te')} ) ~ d\te'\quad
\label{zmikR}	
\\[2pt]
f_{L}(\te)&=&-i \frac{r}{2} e^{-\te} - i   \pi \alpha' 
\nn \\
&& + \int_ {{\cal C}_2}\!\phi(\te-\te')
\ln(1+e^{f_{L}(\te')}) ~ d\te' 
- \int_{{\cal C}_1}\! \phi(\te-\te')
\ln(1+e^{-f_{L}(\te')} ) ~ d\te'  
\nn \\
&& + \int_{{\cal C}_2}\!\chi(\te-\te')
\ln(1+e^{-f_{R}(\te')}) ~ d\te' 
- \int_ {{\cal C}_1}\! \chi(\te-\te')
\ln(1+e^{f_{R}(\te')} ) ~ d\te'\quad
\label{zmikL}	
\eea
and replace the expression (\ref{mnliec}) for the effective central charge
with
\bea
c_{\rm eff}(r)&=& \frac{3 i r }{2 \pi^2} \biggl[ \int_{{\cal
C}_1}\!\!e^{\te} \ln(1{+}e^{f_R(\te)})\,d\te
-\int_{{\cal
C}_2}\!\!e^{\te} \ln(1{+}e^{-f_R(\te)})\,d\te
\nn\\[2pt]
&&{}~~~{}+\int_{{\cal
C}_2}\!\!e^{-\te} \ln(1{+}e^{f_L(\te)})\,d\te
-\int_{{\cal
C}_1}\!\!e^{-\te} \ln(1{+}e^{-f_L(\te)})\,d\te
 \biggr]~.
 \label{zmnliec}
\eea
As before, the contours 
${\cal C}_1$ and ${\cal C}_2$ run from $-\infty$ to $+\infty$
just above and just below the real axis, 
and $r=\MR$. However,
since the flows are massless $M$ no longer has a direct
interpretation as the mass of an asymptotic state, but rather sets the
crossover scale.

In the far infrared, $r\to\infty$ and the two
equations (\ref{zmikR}) and (\ref{zmikL}) decouple.
The result is two copies of the UV limit of (\ref{mnlie}),
with the kernel $\varphi(\theta)$ substituted by $\phi(\theta)$.
Since the IR destination is to be the model at $\xi-1$, we take 
$\phi(\te)$ to be the 
massive kernel (\ref{krnl}) with the substitution $\xi \to \xi-1$:
\eq
\phi(\te)=-\int_{-\infty}^{\infty}
\frac{e^{i k\te} \sinh( k \fr{\pi}{3}) \cosh( \fr{\pi}{6} k (3{-}2 \xi))} 
{\cosh( \fr{\pi}{2} k) \sinh( k \fr{\pi}{3}(\xi{-}1) )}%
\frac{d k}{2\pi}~.
\label{nkrnl}
\en

To find the other kernel, $\chi(\te)$, we consider the UV behaviour of
the new system.  This should coincide with the UV limit
of the massive system (\ref{mnlie}).
As $r\to 0$, the solution to (\ref{zmikR}), (\ref{zmikL})
splits in the standard way into a pair of kink systems, centered at
$\te=\pm\ln(1/r)$. We focus on one of them by replacing $\te$ by
$\te-\ln(1/r)$, and keeping this variable finite as the limit is taken.
The exponential term in (\ref{zmikR}) is replaced by $-\frac{i}{2}e^{\te}$,
while to leading order the exponential in (\ref{zmikL}) can be neglected.
Furthermore, the singularities in $\ln(1+e^{\pm f_L(\theta)})$ 
previously found on the real axis are pushed to $-\infty$, allowing
integration contours 
in any integral involving $f_L$ 
to be moved across the real $\te$ axis at will. It will be
convenient to shift ${\cal C}_2$ down to coincide with ${\cal C}_1$ in all
such integrals, and to
take ${\cal C}_1$ to be the line $\Im m\,\te=-\eta$, with $\eta$ a
suitably-small real number.

For $f_{R}(\te)$, the singularities do not disappear but using the reality
properties of $f_R$ we can at least rewrite the integrations above and below 
the axis in terms 
of the imaginary part of a single integration along ${\cal C}_1$. The final
equation is
\bea
f_{R}(\te)&=& - \frac{i}{2}e^{\te}  + i   \pi \alpha' 
+ 2i  \int_{-\infty}^{\infty}  
\phi(\te-(\te'-i \eta))
\Im m (\ln(1+e^{f_{R}(\te'-i \eta)})) ~ d\te' 
\nn\\[2pt]
&&\qquad\qquad{}-   \int_{-\infty}^{\infty} \chi(\te-(\te'-i \eta))
f_{L}(\te'-i \eta) ~ d\te' 
\label{ftr}
\\[2pt]     
f_{L}(\te)&=& - i   \pi \alpha' +\int_{-\infty}^{\infty} 
\phi(\te-(\te'+ i \eta)) f_{L}(\te'+i \eta)~ d\te' 
\nn\\[2pt]
&&\qquad\qquad{}- 2 i  \int_{-\infty}^{\infty} \chi(\te-(\te'-i \eta))
\Im m (\ln(1+e^{f_{R}(\te'-i \eta)})) ~ d\te' ~.
\label{ftl}
\eea
Taking the Fourier transform of (\ref{ftl}) yields
\eq
(1-\wt{\phi}(k))\,\wt{f}_{L}(k) = - 2 i\, \pi^2 {\alpha}'\delta(k)
 - 2 \, i \, \wt{\chi}(k) \, \wt{L}_{R}(k)
\label{ftll}
\en
where $\wt{L}_{R}(k)$  denotes the transform of 
$\Im m (\ln(1+e^{f_{R}(\te'-i \eta)}))$ and our convention for 
the Fourier transform of $f(\theta)$ is
\eq
\FT [ f(\te) ]   = \int_{-\infty} ^{\infty} f(\te) \, e^{-i \, 
\te \, k} ~ d\te   = \wt{f}(k)
\en
with corresponding inverse
\eq
\FT^{-1} [ \wt{f}(k) ]    = \frac{1}{2 \, \pi} \, 
\int_{-\infty} ^{\infty} \wt{f}(k)%
\, e^{i \, \te \, k} ~ dk =  f(\te)~.
\en
Inserting (\ref{ftll}) into the F.T. of (\ref{ftr}) (first taking
$ -\frac{i}{2} e^{\te} $ to the left hand side to ensure the
 existence of the transform), we obtain
\eq
\FT[ f_{R}(\te) +  \frac{i}{2} e^{\te} ]    =   
2 i \, \pi^2 \alpha' \, \delta(k) \,
\biggl (1+\frac{\wt{\chi}(k)}{1-\wt{\phi}(k)}\biggr )
- 2 \, i \, \wt{L}_{R}(k)\biggl (\wt{\phi}(k)+\frac{\wt{\chi}(k)^2}
{1-\wt{\phi}(k)}\biggr ) ~.
\label{fr}
\en 
This can be compared with the Fourier transform of the kink limit of
(\ref{mnlie}), which is
\eq
\FT[ f(\te) -  \frac{i}{2} e^{\te} ]    =   2i \, \pi^2 \alpha \, \delta(k)
- 2 \, i \, \wt{L}_{R}(k) \, \wt{\varphi}(k) ~.
\label{fmass}
\en 
The two will match if
\eq
\wt{\varphi}(k) = \wt{\phi}(k)+\frac{\wt{\chi}(k)^2}{1
-\wt{\phi}(k)}
\label{chi}
\en
and 
\eq
\alpha =\alpha'\biggl (1+\frac{ \wt{\chi}(0)}{1-\wt{\phi}(0) }\biggr )  ~.
\label{alp}
\en
After some elementary manipulations we can
invert the Fourier transform of $\chi$, resulting in 
\eq
\chi(\te)={-}\int_{-\infty}^{\infty}%
\frac{e^{i k\te}\sinh(\fract{ \pi k}{3} ) 
\cosh(\fract{\pi k}{6} )}%
{\cosh(\fract{\pi k}{2})  \sinh(\fract{\pi k }{3}(\xi-1))  }%
\frac{d k}{2\pi}
\label{chikrnl}
\en
and, from (\ref{alp}),
\eq
\alpha' = \alpha \frac{\xi}{(\xi-1)}~.
\label{alpha}
\en

Since (\ref{chi}) involves only $\chi(k)^2$ we 
have made a choice of sign when taking the square root.
We took the negative sign since only then do the values taken by $c_{\rm
eff}(\infty)$ when flowing from a minimal model $\CM_{p,q}$
always assume the form $1-6/p'q'$, as has to be the case
if the infrared destination of the flow is to be another minimal model.
It is possible that the flows found for the other sign choice also have an
interpretation in some other context, but we will not explore this
question here.
\resection{The flows}
\label{flows}
By construction, the new massless system exhibits the same ultraviolet
central charge as the corresponding massive system:
 \eq
c_{\rm eff}(0)= 1-\frac{3 \xi}{(\xi{+}1)}\alpha^2.
\label{ceffzero}
\en
Furthermore, the fact that the kink limit of the new system can be
mapped exactly onto that of the massive system
makes it natural
to suppose that the recipe for choosing $\xi$
and $\alpha$ given by
(\ref{dt12p}), (\ref{dt21p}) and (\ref{dt15p}) 
also holds for the massless perturbations.
We take this as a working hypothesis for now; in the
next section it will be subject to some more detailed checks.
However, one immediate difficulty should be mentioned: 
with our convention that the coprime pair $(p,q)$ specifying
$\CM_{p,q}$ has $p<q$, the value assigned to $\xi$ by (\ref{dt12p})
for $\phi_{12}$ perturbations is less than $1$. 
The same is true for $\phi_{15}$ perturbations of $\CM_{p,q}$ whenever
$q>4p$. This is not a problem
for the massive perturbations, but the massless
kernel $\phi(\te)$ has a pole at $\te=\frac{2}{3}(\xi{-}1)\pi i$, 
which crosses the
real $\te$ axis as $\xi$ dips below $1$. In addition,
the formula (\ref{alpha}) clearly has a pole at $\xi{=}1$.
All of this leads to technical
complications, and, while they could perhaps be overcome by
analytic continuation, we will leave the issue to one side for now.
In the rest of this paper we will focus solely on the $\phi_{21}$ and
$\phi_{15}$ perturbations, found via equations (\ref{dt21p}) and
(\ref{dt15p}) respectively, and restrict the $\phi_{15}$ perturbations to 
models $\CM_{p,q}$ with $q<4p$.

In the infrared, the massless system mimics a massive system with $\xi$
replaced by $\xi{-}1$ and $\alpha$ replaced
by $\alpha'=\alpha\,\xi/(\xi{-}1)$, and so
\eq
c_{\rm eff}(\infty)
= 1-\frac{3 (\xi{-}1)}{\xi}(\alpha')^2
= 1-\frac{3 \xi}{(\xi{-}1)}\alpha^2.
\label{ceffinf}
\en

Consider first the perturbation of $\CM_{p,q}$ by $\phi_{21}$. 
For it to be relevant, we must impose that
$p>q/2$. Substituting the requisite values of $\xi$ and $\alpha$
(taken from (\ref{dt21p})) into (\ref{ceffzero}) and (\ref{ceffinf})
we find:
\eq
\CM_{p,q}+\phi_{21}~:~~~p>q/2~,~~~~ 
c_{\rm eff}(0)=1-\frac{6}{pq}~~,~~~~
c_{\rm eff}(\infty)=1-\frac{6}{(q{-}p)q}~.
\label{p21r}
\en
Similarly, for the $\phi_{15}$ perturbations we have
\eq
\CM_{p,q}+\phi_{15}~:~~~p<q/2~,~~~~ 
c_{\rm eff}(0)=1-\frac{6}{pq}~~,~~~~
c_{\rm eff}(\infty)=1-\frac{6}{p(4p{-}q)}~.
\label{p15r}
\en
As promised at the end of the last section, the infrared
limiting values of the
effective central charges are all consistent with the 
destinations of the flows being minimal models%
\footnote{had the opposite sign choice for $\chi(\theta)$
been taken in (\ref{chikrnl}), the values of $c_{\rm eff}(\infty)$ 
for the $\phi_{21}$ and $\phi_{15}$ perturbations
would have been $1-6(q{-}p)/(qp^2)$ and $1-6(4p{-}q)/(pq^2)$
respectively}.
Unfortunately, a knowledge of $c_{\rm eff}$ alone
is generally not enough to identify
a nonunitary minimal model uniquely, so the destinations cannot be
completely pinned down by this information.
Moreover, there are cases in which even the  knowledge of 
the full $c_{\rm eff}(r)$
is not sufficient to identify the model or models involved.
The possibility of such an ambiguity was first noted, for
massive perturbations, in~\cite{KTW}. It goes by the name of the
`type II conjecture', and explicitly it equates the effective central
charges of the following models:
\eq
\CM_{p,q}+\phi_{15}
{~~}\leftrightarrow{~~}
\CM_{q/2,2p}+\phi_{21}
{~~}\leftrightarrow{~~}
\CM_{2p,q/2}+\phi_{12}
\qquad (p=2n{+}1,~q=2m)~.
\label{typetwo}
\en
The second equivalence is just a question of labelling, but the first
is much less trivial. However, 
all follow from
the massive NLIE (\ref{mnlie}) and the recipe
(\ref{dt12p})--(\ref{dt15p}), assuming the correctness of these
equations (see~\cite{DT1}).
Clearly the massless theories, whose NLIEs are parametrised according
to essentially the same recipe, will suffer similar ambiguities; 
some examples will be encountered later.

In spite of these provisos, the forms taken by
the results (\ref{p21r}) and (\ref{p15r}) make
for some obvious conjectures, and these will receive further
support, both analytical and numerical, later in the paper. 
Such checks will never resolve type II ambiguities, so for these
cases the conjectures are better motivated by the belief that the
pattern observed in other cases should hold in complete generality.
Taking into account the `$p<q$' labeling convention for a minimal
model $\CM_{p,q}$, the $\phi_{15}$ case splits into two and
we have the following predictions for massless
perturbations of general minimal models:
\bea
\CM_{p,q}+\phi_{21} &\to&  \CM_{q-p,q} \quad \ (\ \ p<q<2p)~, 
\label{fl1}\\[3pt]
\CM_{p,q}+\phi_{15} &\to&  \CM_{p,4p-q}\quad (2p<q<3p)~,
\label{fl2}\\[3pt]
\CM_{p,q}+\phi_{15} &\to& \CM_{4p-q,p}\quad  (3p<q<4p)~.
\label{fl}
\eea
There is only a single flow on this list
from any given minimal model: 
when $\phi_{21}$ is
relevant, $\phi_{15}$ is irrelevant, and vice versa.
Note also that the case (\ref{fl}) can only occur as the
last member of a sequence:
for $3p<q<4p$, $\xi$ lies between $1$ and $2$ and the 
next step would therefore involve a value of $\xi$ less than $1$, 
and we have already decided to exclude such cases.

\begin{figure}[ht]
\begin{center}
\resizebox{.5\linewidth}{!}%
{\includegraphics{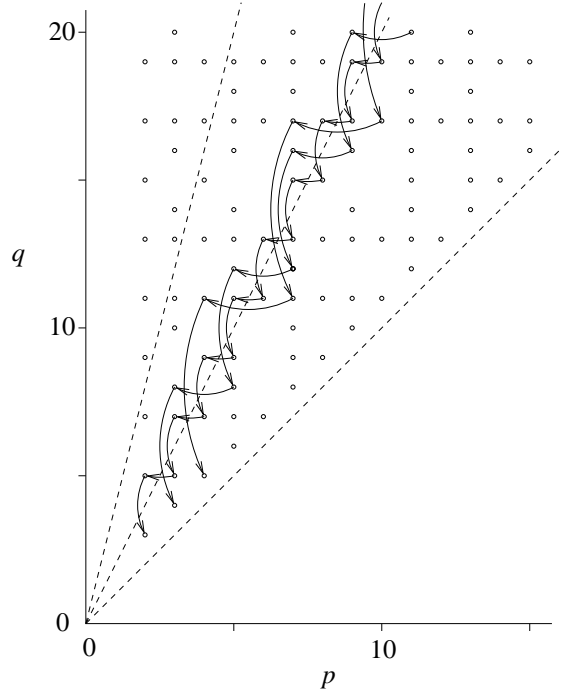}}
\vskip 5pt
\caption{ \protect{\footnotesize 
The grid of minimal models $\CM_{p,q}$ and
a selection of the predicted flows, showing the unique
sequences with $|I|=1$ and $|I|=2$, and one of the two sequences with
$|I|=3$. Also shown are the lines $q{=}p$, $q{=}2p$ and $q{=}4p$.}}
\label{figgrid}
\end{center}
\end{figure}

The previously-known flows (\ref{m21flow}) and (\ref{m15flow}) are
reproduced
on setting $q=2p-1$ in (\ref{fl1}) and $q=2p+1$ in (\ref{fl2}).
In order to understand the more general pattern,
it is convenient to define an  
an `index' $I=2p-q$, and to rephrase (\ref{fl1}) and (\ref{fl2}) as
\bea
\CM_{p,2p-I} +\phi_{21} &\to& \CM_{p-I,2p-I} ~,
\quad(\xi,\alpha')=(\fract{2p}{I}{-}1,\fract{1}{p{-}I})~,
\\[3pt]
\CM_{p,2p+I} +\phi_{15} &\to& \CM_{p,2p-I} ~,
\quad~~~(\xi,\alpha')=(\fract{2p}{I},\fract{2}{2p{-}I})~.
\eea
For reference we have included the values of $\xi$ and $\alpha'$ 
that should be used in the NLIE in order to dial up the corresponding
flow.
In all cases $I({\rm IR}) = - I({\rm UV})$; the flows 
(\ref{m21flow}) and (\ref{m15flow}) 
make up the unique sequence with
$|I|=1$. For $|I|>1$, there may be more than one sequence, with 
flows sharing the same value of $|I|$
interlacing each other. This is reminiscent of the 
generalised staircase models studied in
\cite{DRg}, but is a little more complicated owing to the
constraint that the pair of integers labelling a
minimal model must always be coprime. The number of different
sequences with index $\pm I$ is therefore
given by the Euler $\varphi$-function
$\varphi(|I|)$, equal to the number of integers less than $|I|$ which
are coprime to $|I|$ (so for $n=1\dots 6$, $\varphi(n)=1,1,2,2,4,2$\,).
The index measures the distance from the line
$q=2p$ across which the relevance and irrelevance of the fields
$\phi_{21}$ and $\phi_{15}$ swap over. 
All of this is perhaps best seen pictorially, and in
figure~\ref{figgrid} 
some of the predicted flows
are plotted, superimposed on a grid of the minimal models $\CM_{p,q}$.  
The horizontal arrows of length $|I|$ correspond to $\phi_{21}$
perturbations,  while the vertical arrows, of length $2|I|$, are
$\phi_{15}$ perturbations.

\begin{figure}[ht]
\begin{center}
\resizebox{.6\linewidth}{!}%
{\includegraphics{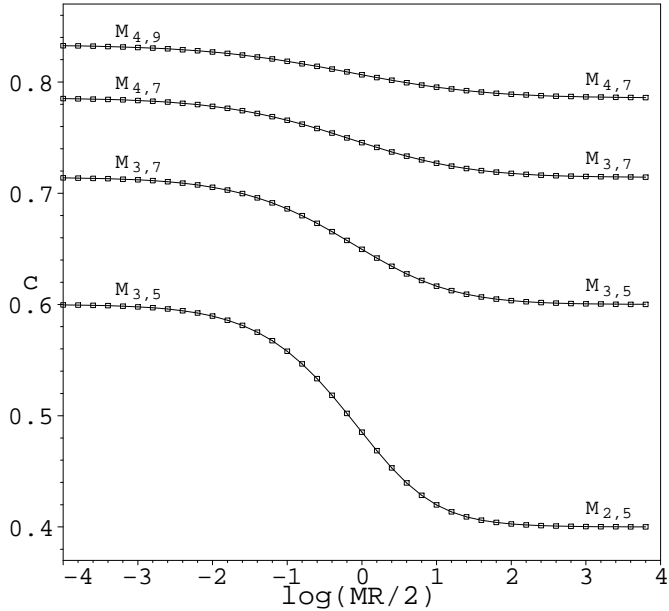}}
\caption{ \protect{\footnotesize 
 TBA (boxes) and NLIE data (solid lines) for the 
`diagonal' sequence with $|I|= 1$. }}
\label{figmart}
\end{center}
\end{figure}

As a first check on our results, we evaluated the effective central
charge numerically for a number of members of the $|I|=1$ series, and made a
comparison
with the results from the massless TBA equations discussed in
\cite{Rac}.
In  \cite{Rac} such equations were written in a `universal'
form of the kind first described in~\cite{Zamy}, but for   
numerical work it is  more convenient to write the equations in the
following way:
\eq
\ep_a(\theta)= \nu_a(\theta) - \sum_{b=1}^{n} 
(l_{ab}^{(A_n)}-\delta_{ab} ) 
\int_{-\infty}^{\infty} 
\CK (\theta-\theta') \ln(1+e^{-\ep_b(\theta')}) \ d \theta'~~,
\label{tba}
\en
\eq
c_{\rm eff}(r)= {3  \over \pi^2} \sum_{a=1}^{n}\int_{-\infty}^{\infty}
 \nu_a(\theta)  \ln(1+e^{-\ep_a(\theta)}) \ d \theta 
\label{tbac}
\en
where 
$n \equiv \xi-3$ is  integer, $l^{(A_n)}_{ab}$ is the incidence matrix of 
the $A_n$ Dynkin diagram,
\eq
\nu_a(\theta)= {r \over 2} (e^{\theta} \delta_{a,1} +
e^{-\theta} \delta_{a,n})~~,
\label{nus}
\en
and
\eq
\CK(\theta)= {\sqrt{3} \over \pi} {\sinh 2 \theta \over \sinh 3 \theta }~~.
\en 
The agreement between the NLIE and the TBA is extremely good,
and is illustrated in table~\ref{tabmart} and figure~\ref{figmart}.
%
\begin{table}[htb]
\begin{center}
\vskip 3mm
\begin{tabular}{c  c  c l l  } 
\hline
Model & \rule[-0.2cm]{0cm}{2mm} $(\xi,\alpha')$~&~~~$r$~~~&~~~~~TBA  
~ &~~~~~NLIE
\\ \hline  
\rule[0.2cm]{0cm}{2mm} 
$\CM_{3,7}
+\phi_{15}
$ & ~($6,\frac{2}{5}$)     
     & 0.01  &  0.71421869185983  &  0.71421869185981 \\
  &   & 0.02   &  0.71408433022934 &  0.71408433022935   \\ \cline{1-5}
\rule[0.2cm]{0cm}{2mm} 
$\CM_{3,5}
+\phi_{21}
$ & ~($5,\frac{1}{2}$)     
     & 0.01  & 0.59996121217228 & 0.59996121217225    \\
  &  & 0.02   &   0.59986582853966  &    0.59986582853967  \\ \cline{1-5}
\end{tabular}
\caption{  \protect{\footnotesize
Comparison of 
$c_{\rm eff}$ 
calculated using the TBA equations and the NLIE.}}
\label{tabmart}
\end{center}
\vspace{-0.2cm}
\end{table}

\begin{figure}[ht]
\begin{center}
\resizebox{.6\linewidth}{!}%
{\includegraphics{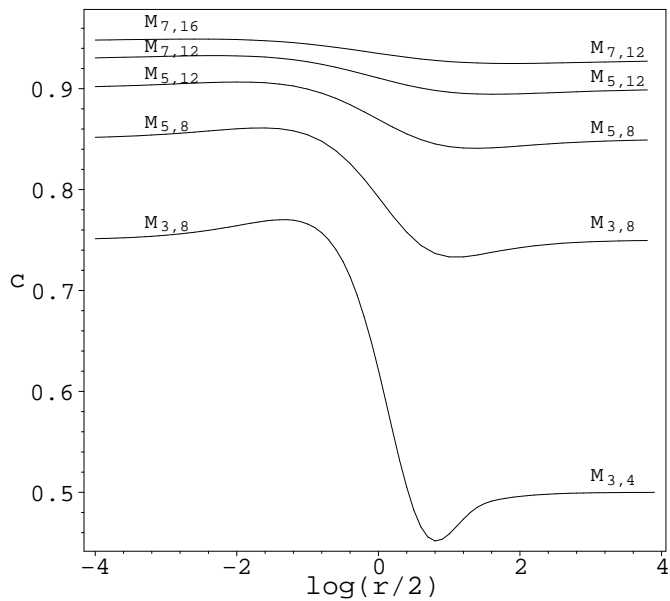}}
\caption{\protect{\footnotesize{The sequence of nonunitary flows with
$|I|=2$.}}}
\label{figieq2}
\end{center}
\end{figure}

\begin{figure}[ht]
\begin{center}
\resizebox{.6\linewidth}{!}%
{\includegraphics{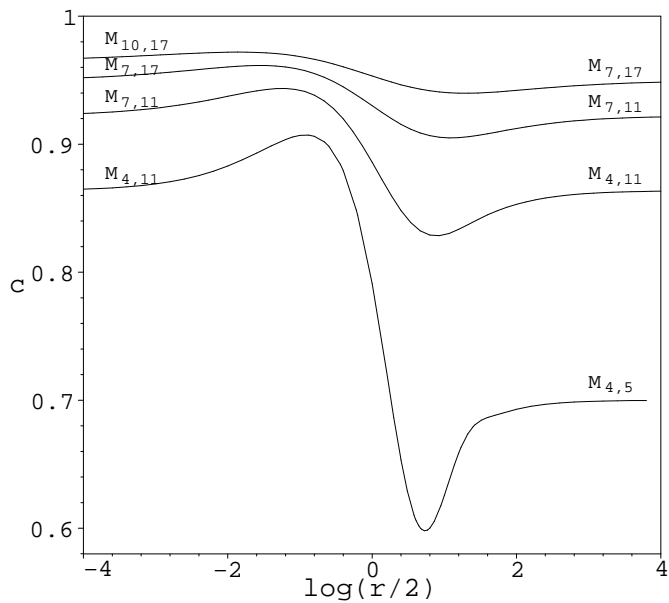}}
\caption{\protect{\footnotesize{
One of the two sequences with $|I|= 3$.
}}}
\label{figieq3}
\end{center}
\end{figure}

Satisfied that our NLIE correctly matches the previously known 
flows, we can turn to the new families with index $|I| > 1$. 
Figure~\ref{figieq2}
shows the (unique) series with $|I|=2$, and 
figure \ref{figieq3} one of the two
possible series with $|I|=3$.
For all of these flows
 the effective central charge initially increases from 
its UV value, 
oscillates as the system size is increased, before finally flowing
to the predicted IR fixed point.  This perhaps-surprising behaviour
is in contrast to the $|I|=1$ 
series where all of the flows are monotonic. We will return to this
point at the end of section~4.  

Note that no TBA equations are known for the massless flows 
with $|I|>1$. This puzzle can be
highlighted by looking at some cases where the corresponding 
massive TBA system {\em can} be conjectured.

\begin{figure}[ht]%
\begin{center}
\resizebox{.6\linewidth}{!}%
{\includegraphics{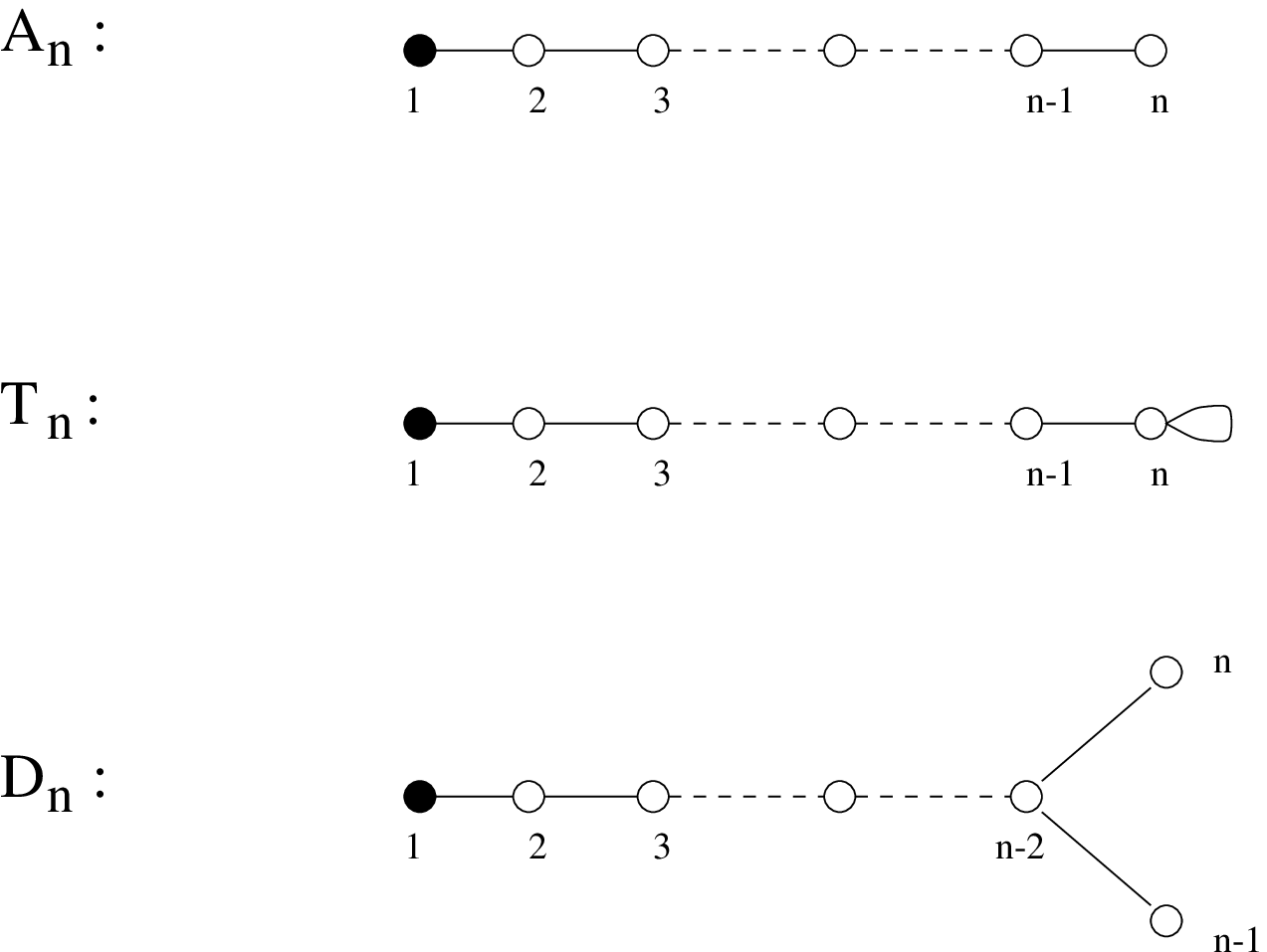}}
\caption{ \protect{\footnotesize 
The $A_n$, $T_n$ and $D_n$ graphs, indicating in each case
the node to which the driving term 
$r \cosh \theta$ in the massive TBA system should be attached.}}
\label{figtba}
\end{center}
\end{figure}

We consider three families of massive systems
found  in the `ADET' class of models 
classified in~\cite{RTV1}.
The first  set is obtained 
from (\ref{tba})
simply by
replacing the driving term 
(\ref{nus}) by
\eq
\nu_a(\theta)=  r  \delta_{a,1} \cosh \theta~.
\label{nus1}
\en
This gives the massive flows corresponding to the 
$|I|=1$ massless flows. For the second set, we
continue to use the driving term (\ref{nus1}), but
replace   $l^{(A_n)}_{ab}$ 
with $l^{(T_n)}_{ab}$
(the incidence matrix of the `tadpole' graph
$T_n=A_{2n}/\Zt$), 
letting $1$ label the node furthest from the `tadpole' node
(see figure \ref{figtba}).
In~\cite{MelzerSUSY} 
these systems were identified 
with the models $\CM_{n+2,2n+2}+\phi_{21}$  
for $n$ odd and  $\CM_{n+1,2n+4}+\phi_{15}$ for $n$ even. 
These are precisely the models in the  $|I|=2$ series.
The massless NLIE predicts, in addition to these previously-known
massive flows, the existence of interpolating flows ($n \rightarrow n-1$) 
within this family.   

The final family of massive TBA equations
is obtained by replacing the $A_n$ incidence
matrix with  the $D_n$ one. For these cases the type II conjecture
mentioned above plays a r\^ole, and
for each TBA system there
are two possible identifications:
\eq
A^{+}_n \equiv \CM_{2n-1,2+4n} +\phi_{15}\quad {\rm and}\quad
A^{-}_n \equiv \CM_{1+2n,4n-2} +\phi_{21}~.
\en
The  sets $A^{\pm}_n$  correspond, in the ultraviolet, 
to the two series of models with index $|I|=4$.
Turning to the massless flows implied by the NLIE, there is
a further ambiguity in the infrared destinations.
However, the general pattern of (\ref{fl1}), (\ref{fl2}) 
and (\ref{flow21})--(\ref{flowvia51}) suggests
\eq
A^{\pm}_n \rightarrow A^{\mp}_{n-1}~~. 
\label{flowsII}
\en

We have checked the low-lying members 
of each family of massive TBA equations
against the massive NLIE (\ref{mnlie}), 
finding agreement to our numerical accuracy (about 14 digits) in each
case.
However, returning to the
question of finding massless versions of these
equations, we observe a key difference
between the first family and the other two: 
as is clear from figure~\ref{figtba},
the graphs for the
$|I|=1$ systems have a $\Zt$ symmetry which is
in general absent from those for $|I|=2$ or $4$ (the single exception 
occurs when $|I|=4$, $n=4$). 
The trick that allowed us to move between massive and massless
systems for $|I|=1$ by 
swapping 
the $\nu$'s defined in (\ref{nus1}) for
those defined in (\ref{nus}) relied completely on this symmetry.
This `$\Zt$ trick' was first employed in \cite{Zc}; 
more generally, the nodes `$1$' and `$n$' could be replaced by any pair 
of nodes related by a $\Zt$ graph symmetry.
To the best of our knowledge, all massless TBA systems that have been
discovered to date are related to massive systems in
essentially this way.
Thus if the $T_n$ and $D_n$ TBA systems do have associated 
massless versions, they
are likely to
be of a somewhat
different nature to all previously-encountered examples.

The $|I|=4$, $n=4$ system
might appear to offer a counterexample, since the exceptional symmetry of 
its graph does allow a massless version of the massive TBA to be constructed.
However this equation turns out not to reproduce the massless NLIE.
For example, for the massless TBA $c_{\rm eff}(\infty)=5/7$, which
identifies the IR destination of the flow as $\CM_{3,7}$ rather than the
$\CM_{7,10}$ or $\CM_{5,14}$ found by the NLIE.
Furthermore, the short-distance expansion of 
$c_{\rm eff}(r)$ turns out to be
inconsistent with the massless TBA flow being related to the massive one by
an analytic continuation of the coupling $\lambda$
-- rather, it is the flow produced by
the massless NLIE which has this property. This failure of the $\Zt$
trick to produce the analytically-continued massless flow can
be put into a more general context. Recall that TBA equations have associated
Y-systems, and that these entail a periodicity 
$Y_a(\theta)=Y_a(\theta+iP)$ for certain functions $Y_a(\theta)$, where $P$
depends on the particular Y-system (see~\cite{Zamy} and also \cite{RTV1}).
Suppose that a diagram symmetry relates nodes $a$ and $\tilde a$ for this
system. 
Then it can be argued that the associated
massive and massless TBA equations
will be related by analytic continuation
if $Y_a(\theta)=Y_{\tilde a}(\theta+iP/2)$, and not otherwise.
In particular, for systems
related to the ADET diagrams, such a property holds if and only if
the nodes $a$ and $\tilde a$ are 
`conjugate', where conjugation acts on TBA diagrams in
the same way as charge conjugation acts on the particles in an
affine Toda field theory \cite{BCDS,D}.
For the $D_n$ diagrams, the fork nodes are related by charge 
conjugation for $n$ odd, but not for $n$ even. This thus matches 
the observation above that the massive and massless $D_4$-related 
systems are not related by continuation in $\lambda$; it also helps
to understand from the TBA point of view
why the models $H_N^{(\pi)}$ of \cite{FZa} are
related to $H^{(0)}_N$ by analytic continuation for $N$ odd, but not
for $N$ even. 

We conclude this section with some further observations.
Following~\cite{MelzerSUSY} we  note  that two  models
in the $|I|=2$ set possess  $N{=}1$ 
supersymmetry in the ultraviolet. In the notation
of~\cite{MelzerSUSY} we have
\eq
S\CM_{2,8}+\hat{\phi}_{13}^{top} \simeq \CM_{3,8}+\phi_{15}~,
\en
and
\eq
S\CM_{3,7}+\hat{\phi}_{15}^{top} \simeq \CM_{7,12}+\phi_{21}~.
\en
The operators 
$\hat{\phi}_{13}^{top}$
and
$\hat{\phi}_{15}^{top}$
are SUSY-preserving, 
so the massless flows 
$\CM_{3,8}\to\CM_{3,4}$ 
and
$\CM_{7,12}\to\CM_{5,12}$ 
should exhibit spontaneous breaking of $N{=}1$ 
SUSY. The situation is reminiscent of the flow  between the tricritical 
Ising and Ising models discussed in~\cite{KMS,Zc}.
In particular, the
flow $\CM_{3,8}\to\CM_{3,4}$
is to a theory of a single
massless free Majorana fermion, just as was the case
in~\cite{KMS,Zc}.
This particle,
in turn, was  identified in~\cite{KMS,Zc} as the  
massless Goldstone fermion.
Finally, we mention that that the massive TBA system
for $\CM_{3,8}+\phi_{15}$ 
was  alternatively obtained  in~\cite{MelzerSUSY}
as a folding of the 
$N{=}2$ supersymmetric model with spontaneously-broken 
$\ZZ_k$  symmetry of~\cite{FI1}, for $k{=}3$.
This suggests that there should also be a (probably
nonunitary) flow from
the `parent' $N{=}2$ supersymmetric theory into 
a massless free fermionic  theory.
Given that  a similar  interpolating
phenomenon is also  present for
$k{=}2$\footnote{for more on 
the $k{=}2$ case, and its connection with the theory of
dense polymers, see~\cite{FSZa,Zd}}
it is natural to suppose that the 
same  behaviour will be exhibited for all of the $\ZZ_k$-related
models of \cite{FI1}.
It would be interesting to check this conjecture using the 
$a_{k-1}^{(1)}$-generalisations of our  NLIE.
\resection{Perturbation theory}
The claims of the last section can be put on a more secure footing
by taking a closer look at the behaviour of the effective central charges
at small
and large $r$.
For this we will need the leading asymptotics of the kernels
$\phi(\theta)$
and $\chi(\te)$
as $\theta\to-\infty$.
That of $\phi$ can be found from the residues of the 
poles in the integrand of (\ref{nkrnl})
at $k=-i$ and $k=-3i/(\xi{-}1)$, and is:
\eq
\phi(\theta)\sim
-\frac{\sqrt{3}\sin(\frac{\pi}{3}\xi)}%
{\pi\sin(\frac{\pi}{3}(\xi{-}1))}\,e^{\theta} 
+
\frac{3\sin(\frac{\pi}{(\xi{-}1)})%
\cos(\frac{\pi}{2(\xi{-}1)})}%
{\pi(\xi{-}1)\cos(\frac{3\pi}{2(\xi{-}1)})}\,e^{3\theta/(\xi{-}1)}
+\dots\,,
\quad \theta\to -\infty~~.
\label{phiasympt}
\en
Similarly,
\eq
\chi(\theta)\sim
-\frac{3}%
{2\pi\sin(\frac{\pi}{3}(\xi{-}1))}\,e^{\theta} 
-
\frac{3\sin(\frac{\pi}{(\xi{-}1)})%
\cos(\frac{\pi}{2(\xi{-}1)})}%
{\pi(\xi{-}1)\cos(\frac{3\pi}{2(\xi{-}1)})}\,e^{3\theta/(\xi{-}1)}
+\dots\,,
\quad \theta\to -\infty~~.
\label{chiasympt}
\en

We first analyse the NLIE as $r\to\infty$, to see whether its
behaviour is compatible with the claimed infrared destinations of the
massless flows.  In spite of the fact that conformal
perturbation theory about the infrared fixed point is not 
renormalisable, there are a number of unambiguous
predictions against which the equation can be checked.
The key ideas are set out in
\cite{ZRSOS,Zc}, and are further discussed in, for example,
\cite{KMspect,FQR,DTT}.

In general, the infrared model will be described by an action of the form
\eq
S=S^*_{\rm IR} +\mu_1\int\psi\,d^2x
+\mu_2\int T\bar T\,d^2x+({\rm further~terms})
\label{iract}
\en
where $S^*_{\rm IR}$ is the action, and $\psi$ one of the
(irrelevant) primary fields, of the infrared conformal field theory. 
$\psi$
might be absent, in which case the only operators attracting the flow to
the IR fixed point would be the descendents of the identity, $T
\bar T$ being the least irrelevant example.
On dimensional grounds, the couplings $\mu_1$ and $\mu_2$ are related
to the single crossover scale $M$ as
\eq
\mu_1=\kappa_1M^{2-2h}~~,~~~\mu_2=\kappa_2M^{-2}~,
\en
with $\kappa_1$ and $\kappa_2$ dimensionless
constants and $h$ the conformal dimension of $\psi$.

Standard methods of perturbed conformal field theory can now be used to
calculate the first corrections to $c_{\rm eff}(\infty)$, with results
that are good at least up to the order at which the `further terms' cut
in. The first perturbing term in (\ref{iract}) yields a 
series in $\mu_1R^{2-2h}=\kappa_1r ^{2-2h}$ if
all powers of $\mu_1$ contribute, or $\mu_1^2R^{4-4h}=\kappa_1^2r^{4-4h}$ 
if only even
powers appear, as happens when $\psi$ is odd under some symmetry of
$S^*_{\rm IR}$. The second term always contributes
a series in $\mu_2R^{-2}=\kappa_2r^{-2}$,
the first two terms of which were found explicitly in~\cite{ZRSOS}.
Putting everything together gives $c_{\rm eff}(r)$ 
the following large-$r$ expansion:
\bea
c_{\rm eff}(r)&\sim&
c_{\rm eff}(\infty)
+(\mbox{a series in $\kappa^{\phantom{2}}_1r^{2-2h}$ or 
$\kappa^{2}_1\,r^{4-4h}$})\nn\\[3pt]
&&\qquad~~~\,
{}-\frac{\pi^3c_{\rm eff}(\infty)^2}{6}\kappa^{\phantom{2}}_2\,r^{-2}
+\frac{\pi^6c_{\rm eff}(\infty)^3}{18}\kappa_2^2\,r^{-4}
+({\rm further~terms})~.\qquad
\label{irexp}
\eea
In contrast to the UV situation to be discussed shortly, 
this series is only expected to be asymptotic, and furthermore there
is very little control over the omitted `further terms'. Nevertheless,
it allows for some useful comparisons with results from the NLIE.

When $r$ is large, the nontrivial behaviours of the functions 
$f_L$ and $f_R$ appearing in (\ref{zmikR}) and (\ref{zmikL}) 
are concentrated in `kink' regions near $\theta=-\ln r$ and 
$\theta=\ln r$ respectively. So long as the contours 
${\cal C}_1$ and ${\cal C}_2$ are kept a finite distance from the real
$\theta$-axis, the functions 
$\ln(1+e^{\pm f_L})$ and
$\ln(1+e^{\pm f_R})$ appearing in these equations are
doubly-exponentially suppressed in the central zone between the kinks,
and so the principal interaction between equations
(\ref{zmikR}) and (\ref{zmikL}), and hence the principal correction
to $c_{\rm eff}(r)$, comes from the exponential tail 
of the kernel function $\chi(\theta)$, given by (\ref{chiasympt}).
Since we are only worrying about the first few terms in the expansion,
and we are free to vary $\xi$ so as to avoid any `resonance' effects, we
can discuss effects of the two terms in (\ref{chiasympt}) separately.
Consider first the second term, decaying as $e^{3\theta/(\xi-1)}$ as
$\theta\to-\infty$. Inserted into (\ref{zmikR}) and considered
iteratively, corrections to $f_R(\theta)$ as
a series in $r^{-6/(\xi{-}1)}$ will be generated. Feeding through into
$c_{\rm eff}(r)$, these corrections
can be matched against the $\kappa_1$ terms in
(\ref{irexp}), allowing $h$, the conformal dimension of the field $\psi$,
to be extracted.
Comparing with the Kac formula $h_{ab}=((bp'{-}aq')^2-(p'{-}q')^2)/(4p'q')$
at the appropriate (IR) values of $p'$ and
$q'$ leads us to conjecture
the following pattern of arriving operators $\psi$:
\bea
\CM_{p,q}+\phi_{21} &\to&  \CM_{q-p,q} \quad \; \, (\ \ p<q<2p)
\quad\hbox{arriving via $\phi_{21}$}~,\label{flow21}\\[3pt]
\CM_{p,q}+\phi_{15} &\to&  \CM_{p,4p-q}\quad (2p<q<3p)
\quad\hbox{\,arriving via $\phi_{15}$}~,\label{flowvia15}\\[3pt]
\CM_{p,q}+\phi_{15} &\to& \CM_{4p-q,p}\quad (3p<q<4p)
\quad\hbox{\,arriving via $\phi_{51}$}~.\label{flowvia51}
\eea
In making these identifications, we took account of the fact that
$\phi_{21}$ is an odd operator, while $\phi_{15}$ is even.

Exactly the same line of argument starting from the $e^{\theta}$ term in 
(\ref{chiasympt}) reveals a further series of
corrections to $c_{\rm eff}$ as powers of $r^{-2}$, perfectly adapted to
match the $T\bar T$ terms in (\ref{irexp}). However this time it is
possible to say more, by observing that the effect of this 
part of $\chi(\theta)$ on equations (\ref{zmikR}) and (\ref{zmikL})
can be reabsorbed into a
shift of $r$ by a constant, and furthermore that this constant can
be expressed in terms of $c(r)$. This allows the first few terms of the
iterative expansion to be found exactly. The same idea was used 
in \cite{DTT} to extract IR asymptotics from various
massless TBA systems; in the current context we find
\eq
c_{\rm eff}(r)\sim c_{\rm eff}(\infty)
-\frac{\pi c_{\rm eff}(\infty)^2}{\sin\frac{\pi}{3}(\xi{-}1)}r^{-2}
+\frac{2\pi^2 c_{\rm eff}(\infty)^3}{\sin^2\frac{\pi}{3}(\xi{-}1)}r^{-4}
+\dots
\label{TTpred}
\en
Matching $r^{-2}$ terms in (\ref{irexp}) and (\ref{TTpred}) gives
the exact relation
$\kappa_2=6/(\pi^2\sin\frac{\pi}{3}(\xi{-}1))$\,;
the fact that the $r^{-4}$ terms then agree provides a
nontrivial check on the IR behaviour of the massless NLIE. 

We also performed some numerical fits on the IR data. Our results are
summarised in table~\ref{tabirfits}. Accuracy was fairly low, but 
where relevant we have also included the predicted (`exact') values
of coefficients, obtained from equation (\ref{TTpred}).

\begin{table}[htb]
\begin{center}
\vskip 3mm
\begin{tabular}{l l    } 
\hline
{}~Model & ~~~$c_{\rm eff}(r){-}c_{\rm eff}(\infty)$
\rule[-0.2cm]{0cm}{2mm}\\ 
\hline 
\rule[0.4cm]{0cm}{2mm} 
$\CM_{3,5}+\phi_{21}$   &
$  \ \  0.58041577 \, r^{-2}
 +   0.000061 \, r^{-3}  + 1.66 \, r^{-4}   +0.12 \, r^{-\frac{9}{2}}
+\dots$ \\
\rule[0.2cm]{0cm}{2mm} 
Exact   &
$ \ \ 0.58041579 \,r^{-2}+ 1.6844 \, r^{-4}$ 
    \\
\rule[0.4cm]{0cm}{2mm} 
$\CM_{3,7}+\phi_{15}$  &
$ \  \
  1.30591\,r^{-2}
+  1.65488\,r^{-\frac{ 12}{5}} +\dots$ \\
\rule[0.2cm]{0cm}{2mm} 
Exact & $ \ \ 1.30593 \,r^{-2} $\\
\rule[0.4cm]{0cm}{2mm}
$\CM_{4,7}+\phi_{21}$&
$ 
  \ \ 0.51029\,\ln (r) \,r^{-2} -    0.68683\,r^{-2} +\dots
$ \\
\rule[0.2cm]{0cm}{2mm} 
Exact &
$ \ \ 0.51020 \, \ln (r) \,r^{-2}$ \\
\rule[0.4cm]{0cm}{2mm}  
$\CM_{7,11}+\phi_{21}$&
$  -  6.8516\,r^{-2} +   8.8213\,r^{-\frac{ 9}{4}} +\dots
$ \\
\rule[0.2cm]{0cm}{2mm} 
Exact &
$ -6.8510 \, r^{-2} $ \\
\rule[0.4cm]{0cm}{2mm} 
$\CM_{5,8}+\phi_{21} $&
$ -    1.1253\,\ln (r)\,r^{-2}+    1.3757\,r^{-2} +\dots
$ \\
\rule[0.2cm]{0cm}{2mm}
Exact  &
$ -1.125 \ \  \ln (r) \, r^{-2}$ \\
\rule[0.4cm]{0cm}{2mm} 
$\CM_{6,17} + \phi_{15} $ &
$ -2.32081 \, r^{-2}
+ 12.85 \, r^{-4}
 -1.527 \, r^{\frac{-30}{7}} 
 +\dots $ \\
\rule[0.2cm]{0cm}{2mm}
Exact  &
$  -2.32082 \,r^{-2} + 12.56 \, r^{-4} $ \\[3pt]
\hline
\end{tabular}
\caption{  \protect{\footnotesize
Comparison of the 
infrared expansion of 
$c_{\rm eff}(r){-}c_{\rm eff}(\infty)$ for various models
against the exact coefficients. }}
\label{tabirfits}
\end{center}
\vspace{-0.2cm}
\end{table}

The situation in the ultraviolet is in many respects much simpler.
Conformal perturbation theory gives direct access to a function 
$c_{\rm pert}(r)$, related to the ground state energy
(\ref{Egdef}) as $E_0(M,R)=-\frac{\pi}{6R}c_{\rm pert}(r)$.
Note that $c_{\rm pert}$ contains a bulk part
which must be subtracted before comparisons are made with the NLIE.
For a theory perturbed by a relevant primary operator
$\phi$ with scaling dimensions $(h_{\sss UV},h_{\sss UV})$\,,
$c_{\rm pert}$ has the expansion~\cite{ZTBA,KM}
\eq
c_{\rm pert}(r)=c(0)+ \sum_{n=1}^\infty C_n (\lambda R^y)^n
\label{cpert}
\en
and, in contrast to the situation in the IR, the series
is expected to have a finite radius of convergence.
Here $y=2-2h_{\sss UV}$, $\lambda$ is the coupling, and
the coefficients $C_n$ 
are given in terms of the connected correlation functions
of the perturbing field on the plane as
\eq
C_n= \frac{12\,(-1)^n}{n!\,(2\pi)^{yn-1}} 
\,\int \prod_{j=2}^{n}\frac{d^2z_j}{|z_j|^y}\,
\langle V(0)\phi(1,1)\phi(z_2,\bar z_2)\dots\phi(z_n,\bar z_n) 
V(\infty)\rangle^{\phantom{1}}_{\sss C}~,
\en
where $V$ creates the CFT ground state on the
cylinder. This is the 
state with lowest conformal dimension, and for a general minimal model
$\CM_{p,q}$ it corresponds to the field $\phi_0\equiv
\phi_{ab}$ with $a$ and $b$ 
integers satisfying $bp-aq=1$. (Only for the unitary models
$\CM_{p,p{+}1}$ does it coincide with the conformal vacuum $\phi_{11}$.)

For the perturbing operator, we
will be interested in $\phi=\phi_{15}$ and
$\phi=\phi_{21}$. As mentioned above,
$\phi_{21}$ is odd, so
only the even coefficients $C_{2n}$ are nonzero for this
case. Since
$\lambda$ must be related to the crossover scale $M$ as
\eq
\lambda=\kappa M^{y}
\en
with $\kappa$ a dimensionless constant, 
(\ref{cpert}) is a series in $r^y$ for the $\phi_{15}$
perturbations, and $r^{2y}$ for the $\phi_{21}$ perturbations.

The effective central charge calculated using the NLIE is expected to 
expand as 
\eq
c_{\rm eff}(r)=c_{\rm eff}(0) + B(r) + \sum_{n=1}^{\infty}c_n 
r^{yn}\,
\label{ceq}
\en
for some value of $y$.
Periodicity arguments suggest that this will be a series in
$r^{6/(1+\xi)}$, and if $\xi$ is chosen according to
(\ref{dt21p}) or (\ref{dt15p}) then
$y=2-2h_{\sss UV}$ and
the perturbative expansion (\ref{cpert}) is matched,
with $h_{\sss UV}$ the conformal dimension of
either $\phi_{21}$ or $\phi_{15}$, and all odd terms zero for
the $\phi_{21}$ case~\cite{DT1}.
The irregular bulk term
$B(r)$ 
must be subtracted before $c_{\rm pert}$ can be compared
with $c_{\rm eff}$.
Fortunately, it can be obtained exactly from the NLIE,
using a small generalisation
of arguments used in \cite{ZTBA} and \cite{DDV}.
The term we need
is given by the behaviour as $r\to 0$ of
\eq
2\,\frac{3 ir}{2 \pi^2} \biggl[ \int_{{\cal
C}_2\atop >0}\!\!e^{-\te}\frac{d}{d\te}\ln(1{+}e^{f_L(\te)})\,d\te
-\int_{{\cal
C}_1\atop >0}\!\!e^{-\te}\frac{d}{d\te}\ln(1{+}e^{-f_L(\te)})\,d\te
 \biggr]~,
 \label{inii}
\en
where the `$>0$' indicates that only those parts of the contours 
${\cal C}_1$ and ${\cal C}_2$ with positive real part should be taken, and
the symmetry between $f_L$ and $f_R$ was used to trade the first two
integrals in (\ref{zmnliec}) for the prefactor $2$.
Consider the
$r\to 0$ limit of 
(\ref{zmikR}) and (\ref{zmikL}) 
in the region $0\ll\Re e\,\te\ll\ln(1/r)$\,, where the driving term
$-i\frac{r}{2}e^{-\te}$ in (\ref{zmikL}) can be dropped. Take the
derivative of these equations with respect to $\te$, and extract the
contributions to the convolutions proportional to $e^{\te}$ using
(\ref{phiasympt}) and (\ref{chiasympt}). These should cancel
either against the remaining driving term or
between themselves to ensure that the 
functions $f_L$ and $f_R$ have no such dependency. This leads to the
equations
\bea
i \frac{r}{2}
\!\!&=& \!\!
-\frac{\sqrt{3}\sin(\frac{\pi}{3}\xi)} {\pi\sin(\frac{\pi}{3}(\xi{-}1))}
\biggl[\int_ {{\cal C}_1\atop >0}\!e^{-\te'}\frac{d}{d\te'}
\ln(1{+}e^{f_{R}(\te')}) \, d\te' 
- \int_ {{\cal C}_2\atop >0}\! e^{-\te'}\frac{d}{d\te'}
\ln(1{+}e^{-f_{R}(\te')} ) \, d\te' \biggr] 
\nn \\[2pt]
&&
\!\!\!\!-\frac{3}%
{2\pi\sin(\frac{\pi}{3}(\xi{-}1))}
\biggl[ \int_ {{\cal C}_1\atop >0}\!e^{-\te'}\frac{d}{d\te'}
\ln(1{+}e^{-f_{L}(\te')}) \, d\te' 
- \int_ {{\cal C}_2\atop >0}\! e^{-\te'}\frac{d}{d\te'}
\ln(1{+}e^{f_{L}(\te')} ) \, d\te'\biggr]\,;
\nn\\[2pt]
0\!&=&\!
-\frac{\sqrt{3}\sin(\frac{\pi}{3}\xi)} {\pi\sin(\frac{\pi}{3}(\xi{-}1))}
 \biggl[ \int_ {{\cal C}_2\atop >0}\!e^{-\te'}\frac{d}{d\te'}
\ln(1{+}e^{f_{L}(\te')}) \, d\te' 
- \int_{{\cal C}_1\atop >0}\! e^{-\te'}\frac{d}{d\te'}
\ln(1{+}e^{-f_{L}(\te')} ) \, d\te'\biggr]  
\nn \\[2pt]
&&
\!\!\!\!-\frac{3}%
{2\pi\sin(\frac{\pi}{3}(\xi{-}1))}
\biggl[ \int_{{\cal C}_2\atop >0}\!e^{-\te'}\frac{d}{d\te'}
\ln(1{+}e^{-f_{R}(\te')}) \, d\te' 
- \int_ {{\cal C}_1\atop >0}\! e^{-\te'}\frac{d}{d\te'}
\ln(1{+}e^{f_{R}(\te')} ) \, d\te'\biggr]\,.
\nn\eea
Solving for the integrals needed for the evaluation of
(\ref{inii}), we find
\eq
B_{massless}(r)
=-\frac{3}{\pi}\,\frac{\sin(\frac{\pi}{3}\xi)\sin(\frac{\pi}{3}(\xi{-}1))}%
{\sin(\pi\xi)}\,r^2~.
\label{bulki}
\en
A similar if slightly simpler analysis of the massive equation
(\ref{mnlie}) predicts
\eq
B_{massive}(r)
=-\frac{\sqrt{3} \, \sin( \frac{\pi}{3} \xi ) }
   {2 \, \pi   \sin ( \frac{\pi}{3} (\xi{+}1)) } \,r^2~.
\label{bulk}
\en
It can be checked that this
formula combines 
the results given separately for
$\phi_{12}$, $\phi_{21}$ and $\phi_{15}$ perturbations in
\cite{FT,FLZZ}.  

The formula (\ref{bulki}) has a pole 
when $\xi+1$ is an integer multiple of 3. Since the 
overall result remains finite this infinity 
should cancel 
against one of the  
terms in the regular expansion of $c_{\rm pert}$, leaving
a logarithmic contribution to the final result (see for example
\cite{HF,CM,DTT}).  At this point 
the calculation 
might appear to split
into two cases: 
the term proportional to
$r^2$ in the perturbative expansion
occurs at an order $n$ which depends both on the dimension
and on the `parity' of the perturbing operator:
$n=2/y$ for  $\phi_{15}$, and
$1/y$ for $\phi_{21}$\,.
However $y=6/(\xi{+}1)$ for $\phi_{15}$ and
$3/(\xi{+}1)$ for $\phi_{21}$ \cite{DT1},
so $n=(\xi{+}1)/3$ in both cases and they can
be treated simultaneously. 
The logarithm is found by evaluating
\eq
\lim_{\xi \to 3n-1 } -\frac{{3} \, \sin( \frac{\pi}{3} \xi )%
\sin( \frac{\pi}{3} (\xi{-}1) )}
{\pi   \sin ( \pi \xi) } \,\biggl(r^2-r^{2n\frac{3}{\xi+1}}\biggr)
\en
which yields
\eq
B_{massless}(r)\Bigl|_{\xi=3n-1}
=(-1)^n  \frac{3} {2 \,  \pi^2 \, n }\,r^2\ln r ~.
\qquad (n\in\NN)
\en
Similarly, and
for exactly the same values of $\xi$,  logarithms arise
in the massive perturbations.
In these cases we find
\eq
B_{massive}(r)\Bigl|_{\xi=3n-1}
=\frac{3} {2 \, \pi^2  \, n }\,r^2\ln r ~.
\qquad \qquad \ \ \ (n\in\NN)
\en

This seems to leave the coefficient of $r^2$ term unknown, but 
one piece of information can be extracted, 
which will be used later.  Recall that the logarithm
is linked to a divergence in the perturbative integral defining
the term $C_n$. 
Providing the same regularisation scheme is used
for both flow directions, 
the perturbative contributions should cancel 
upon taking their sum or difference.
The remaining finite contribution can be found directly
from (\ref{bulki}) and (\ref{bulk}):
\eq
\Bigl(B_{massless}(r) \pm B_{massive}(r)\Bigr)\Bigl|_{\xi=3n-1}
= \pm \frac{\sqrt 3}{4 \, \pi}\,r^2
\qquad (n\in\NN)
\label{bulkr2}
\en
where the plus signs should be chosen if $n$ is odd and the minus signs
if $n$ is even.  

Using an iterative method to solve (\ref{zmikR}) and  (\ref{zmikL}),
$c_{\rm eff}(r)$ can be calculated to high accuracy.
After subtracting the 
ultraviolet central charge and the relevant bulk term, the perturbative
coefficients $c_n$ can be estimated via a polynomial fit.  
To make the comparison with CPT, we must also fix the value of
$\kappa$. For the massive systems this was determined exactly in
\cite{FT,FLZZ}, and it turns out, numerically at
least, that the same
relationship  between $\lambda$ and $M$
holds in the massless cases, apart from a factor of either
$-1$ or $i$.
As explained in \cite{Rac}, the relationship between the massive and 
massless perturbations depends on the parity of the perturbing operator.
For $\phi_{15}$\,, the massive behaviour is related to the 
massless by flipping the sign of the coupling constant $\lambda$. This 
has no effect for the odd operator $\phi_{21}$; instead the 
required transformation is $\lambda \to i \lambda$.
The CPT coefficients $C_n$ are the same for
both flow directions so,
provided the mass and the crossover scales $M$ are equal,
we expect
\bea
\phi_{15}:~~~~~~~~~c_n&=&(-1)^n \ \wt{c}_n~, 
\label{creln15}\\
\phi_{21}:~~~~~~~~c_{2n}&=&(-1)^n \ \wt{c}_{2n}~,
\label{creln21}
\eea
where $\wt{c}_n$ denotes the expansion coefficients obtained using
the massive NLIE.
Tables~\ref{tabm411} and \ref{tabm58} present the first few 
perturbative coefficients for the models $\CM_{4,11}$
and $\CM_{5,8}$ perturbed in both massless 
and massive directions.  The relative signs of the coefficients
confirm relations (\ref{creln15}) and (\ref{creln21}). Similar results
were obtained for the models 
$\CM_{3,8}$\,,
$\CM_{5,12}$ and
$\CM_{3,10}$ perturbed by $\phi_{15}$\,, and the models
$\CM_{3,4}$\,,
$\CM_{3,5}$ and
$\CM_{7,11}$ perturbed by $\phi_{21}$. 
These provide strong support for our claim that, modulo the
bulk terms, the results from the
massive and massless integral equations are related by analytic
continuation.
\begin{table}
\begin{center}
\begin{tabular}{ c l l l  }                          \hline 
$n$ &   $~\wt{c}_n$    (massive)           &   $~c_n$     (massless)
& $~\Delta c_n $          \\
\hline  
0 &~19/22           &~19/22          & $\ \ \ \,  ~/$ \\
1 &-0.393148695201  &~0.393148695201 & $\pm 7\cdot 10^{-13}$ \\
2 &~0.0166115829    &~0.0166115834   & $\pm 5\cdot 10^{-10}$ \\
3 &-0.001225836     &~0.001225827    & $\pm 8\cdot 10^{-9}$  \\
4 &-0.00003425      &-0.00003422     & $\pm 2\cdot 10^{-8}$  \\
5 &~0.0000116       &-0.000012       & $\pm 3\cdot 10^{-7}$  \\
\hline 
\end{tabular}
\caption{
\protect{\footnotesize Comparison of massive and massless 
UV coefficients for 
$\CM_{4,11}{+}\phi_{15}$. The last column
reports the difference between the absolute values of the two.} }
\label{tabm411}
\end{center}
\end{table}
\begin{table}
\begin{center}
\begin{tabular}{c l l l  }                          \hline 
$n$ &   $~\wt{c}_n $ (massive)                &   $~c_n$ (massless) 
& $~\Delta c_n$  \\ \hline 
0 &~0.85            &~0.85             & $\ \ \ \, ~/  $ \\
2 &-0.1128538069088 &~0.1128538069085  & $\pm 3\cdot 10^{-13}$ \\
4 &~0.1607667041    &~0.1607667048     & $\pm 7\cdot 10^{-10}$ \\
6 &~0.003799534     &-0.003799535      & $\pm 1\cdot 10^{-9}$  \\
8 &-0.0009586       &-0.0009585         & $\pm 1\cdot 10^{-7}$  \\
10&-0.0001554       &~0.0001554        & $\pm 8\cdot 10^{-8}$  \\
\hline 
\end{tabular}
\caption{
\protect{\footnotesize Comparison of massive and massless 
UV coefficients for 
$\CM_{5,8}+\phi_{21}$. The last column
reports the difference between the absolute values of the two.} }
\label{tabm58}
\end{center}
\end{table}

For the $\phi_{15}$ perturbations, it is also simple to
check the value of $\kappa$ directly, since the first term
in the perturbative expansion (\ref{cpert}) is just given by
\eq
C_1=-12\,(2\pi)^{(1-y)}C_{\phi_0 \, \phi_{15} \, \phi_0}
\label{c1}
\en
where $C_{\phi_0\,\phi_{15}\,\phi_0}$\,,
the operator product coefficient 
between the perturbing operator $\phi_{15}$ 
and the ground state $\phi_0$\,, can be found in \cite{FD}.
Comparing (\ref{cpert}) and (\ref{ceq}) at order $n=1$, we have
\eq
\kappa=\frac{c_1}{C_1}~.
\label{kappa}
\en
Using this formula and the massless NLIE, we estimated
$\kappa^2$ numerically for a number of models.
Table~\ref{tabkap} reports the results, and
compares them with the exact
expression for
$\kappa^2$ for (massive) $\phi_{15}$ perturbations
given in~\cite{FLZZ}:
\eq
\lambda^2=\frac{ 4^2 \, (\xi-1)^2   
\  \gamma(\frac{2+\xi}{2(1+\xi)})
\ \gamma(\frac{5\xi}{2(1+\xi)} )     } 
  {\pi^2 \  (2-3\xi)^2 \ (2-\xi)^2 \ 
\gamma^2(\frac{3+\xi}{1+\xi})  }
\biggl[ \frac{ M \  
\Gamma(\frac{\xi+1}{3}) }
{ \sqrt 3 \  
\Gamma(\frac{1}{3})\  \Gamma(\frac{\xi}{3})}    \biggr] ^{ \frac{12}{\xi+1}}
\en
where $\gamma(x)=\frac{\Gamma(x)}{\Gamma(1-x)}$ and $M$ is the mass
of the lightest kink.\footnote{note the relation was
given in \cite{FLZZ} 
in terms of  $\xi^{\rm FLZZ}=\frac{p}{q-p}$ and $m$, the 
mass of the lightest breather, related to $M$
as $m=2 M \, \sin(\frac{\pi}{3}\xi)$}
The agreement is clearly very good. Note that
$\kappa^2$ is sometimes negative -- these are cases where,
in the normalisations of \cite{FD}, 
$C_{\phi_0\,\phi_{15}\,\phi_0}$ turns out to be purely imaginary.

\begin{table}
\begin{center}
\begin{tabular}{ c l l}   
\hline 
Model  & $~~~~~~~~\kappa_{num}^2$ & $~~~~~~~~\kappa^2 $  \\
\hline 
$\CM_{5,12}$  &~~0.012665147953   &~~0.012665147955 $\dots$  \\
$\CM_{7,16}$  &~~0.0247653386712  &~~0.0247653469711 $\dots$  \\ 
$\CM_{4,11}$  &~-0.06407485530    &~-0.06407485531 $\dots$  \\
$\CM_{7,17}$  &~~0.0092588732986  &~~0.0092588732985 $\dots$ \\
$\CM_{10,23}$ &~~0.02340788397    &~~0.0234078837 $\dots$  \\
$\CM_{3,10}$  &~-0.0188367        &~-0.0188368 $\dots$  \\
\hline
\end{tabular}
\caption{
\protect{\footnotesize 
Comparison of exact and numerical values of $\kappa^2$ for 
$\phi_{15}$ perturbed models.
} }
\label{tabkap}
\end{center}  
\end{table}

Now we return to the behaviour of the flows
(\ref{m21flow}) and (\ref{m15flow}), to mention one reason why,
exceptionally, $c_{\rm eff}(r)$ should be a monotonic function of 
$r$ for these flows.
The behaviour
of a flow is determined at small $r$ by the first nonzero
term in the perturbative expansion. Thus
whether $c_{\rm eff}$ initially increases or decreases will
typically
be determined by
the sign of $C_1\lambda$ ($\phi_{15}$) 
or $C_2\lambda^2$ ($\phi_{21}$).
In particular, if $C_1$ (respectively $C_2$) is nonzero then
for one sign of $\lambda$ (or $\lambda^2$), $c_{\rm eff}$ will 
initially increase, leading to an immediate violation of the 
`$c_{\rm eff}$-theorem', and a 
flow which must be non-monotonic
if $c_{\rm eff}(\infty)$ is to be less than
$c_{\rm eff}(0)$.
But for the model $\CM_{3,5}+\phi_{21}$,
the first coefficient $C_2$ 
was calculated to be zero in \cite{Rac}.
In this case
the asymptotic UV behaviour is instead controlled by $C_4\lambda^4$ 
and this permits $c_{\rm eff}$ to decrease
initially for both (massless and massive) signs of $\lambda^2$.
A similar calculation for 
the other models in the $|I|=1$
series finds 
$C_1$ ($\phi_{15}$) 
or $C_2$ ($\phi_{21}$) to be zero and
thus, as in the first model, 
all of the flows are able to be monotonic.
A numerical fit of the 
data for models higher up in the series confirms these coefficients
are zero within our numerical accuracy. For all other sequences of
models, such vanishings of $C_1$ or $C_2$ do not occur, forcing at
least one of each pair of massive and massless flows to be
non-monotonic. One remaining mystery, out of many,
is why it should always be the
{\em massless} flow which is non-monotonic.

The $|I|=1$ sequence of flows 
was found in \cite{Ma} via an associated staircase model, and the
final step of this staircase
interpolates in the infrared to a massive model with $c_{\rm eff}=0$.
It turns out that the NLIE also reproduces this behaviour:
even though the operator $\phi_{15}$ is not a member of the 
Kac table for $\CM_{2,5}$\,,
(\ref{fl2}) formally predicts a further flow, 
to a theory with $c_{\rm eff}=0$:
$\CM_{2,5} + \phi_{15} \to \CM_{2,3}$.
There is nothing to
stop us using the massless NLIE (\ref{zmikR},\ref{zmikL})
and the $\phi_{15}$ recipe
(\ref{dt15p}) to compute an effective central 
charge for this putative flow.
{}Our numerical results show $c_{\rm eff}$ to have
an exponential behaviour in the infrared, 
as would be expected for a massive, rather than a 
massless, flow.  
Furthermore,
comparing $c_{\rm eff}$ (massless) with that 
of the massive flow $\CM_{2,5}+\phi_{15}$ calculated
using (\ref{mnlie}) and (\ref{dt15p}), 
we find the effective central charges
match exactly.
They also turn out to
coincide with the results
from the more standard massive flow $\CM_{25}+\phi_{12}$, computed
using (\ref{mnlie}) and (\ref{dt12p}). This 
means that the flow is at least physically reasonable, since
$\phi_{12}$ is the single relevant primary 
field in $\CM_{2,5}$. However, it remains a curiosity
that the same flow can be found from three different
nonlinear integral equations.
A sample of our numerical results is shown in table~\ref{tabm25}.
Note that this example shows that the straightforward prediction
of scaling dimensions based on periodicity arguments is not always correct:  
for $\CM_{2,5}+\phi_{15}$ one would have expected a series  
in $r^{6/5}$ for $c_{\rm eff}(r)$, whereas in fact the expansion 
is in powers of $r^{12/5}$.
\begin{table}[htb]
\begin{center}
\vskip 3mm
\begin{tabular}{ l   c  c  c }
\hline 
r & $\CM_{2,5}+\phi_{15}$ (massless) & $\CM_{2,5}+\phi_{15}$ (massive)&
   $\CM_{2,5}+\phi_{12}$ (massive)    \\
\hline 
0.01 & 0.39997512539833 &  0.39997512539839  &   0.39997512539863  \\
0.2 &  0.39254149935036  & 0.39254149935037 &    0.39254149935054  \\
 \hline 
\end{tabular}
\caption{\protect {\footnotesize Comparison
of effective central charge for the three 
a priori different models.
}}
\label{tabm25}
\end{center}
\vspace{-0.2cm}
\end{table}

\begin{figure}[ht]
\centerline{
\resizebox{.6\linewidth}{!}%
{\includegraphics{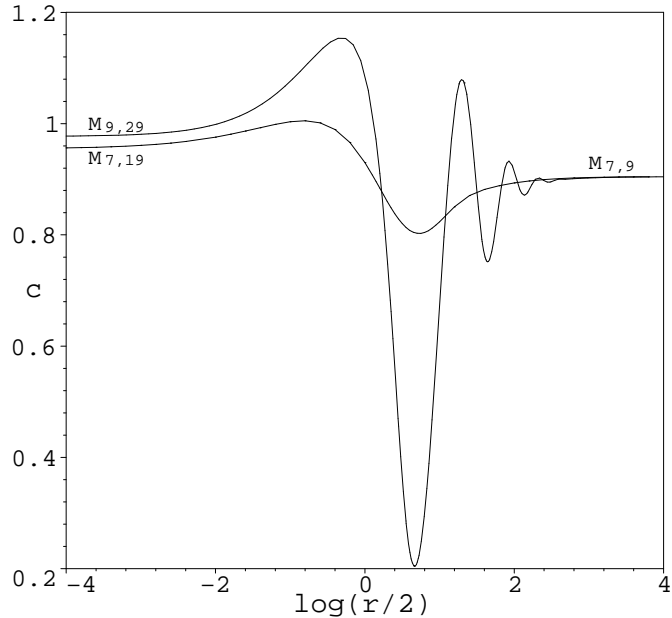}}
}
\caption{\protect{\footnotesize 
Two different flows to the same IR point,
$\CM_{7,9}$\,: $\CM_{9,29} +
\phi_{15}$ attracted via $\phi_{51}$\,, and $\CM_{7,19} 
+\phi_{15}$ attracted via $\phi_{15}$\,\,. 
}}
\label{figtom79}
\end{figure}

If we compare the infrared destinations of
(\ref{fl2}) and (\ref{fl}) we see it is possible for
two different ultraviolet models, both perturbed by $\phi_{15}$,
to  flow to the same 
infrared fixed point,  one attracted via the irrelevant operator
$\phi_{15}$ 
and the other via  $\phi_{51}$. 
Figure~\ref{figtom79} illustrates two such flows.
As was noted in \S3, the sequence attracted by  $\phi_{51}$ 
necessarily stops at this model,
but the other sequence may continue to flow down further.

For some models, the predictions
(\ref{flow21})--(\ref{flowvia51}) must be treated with caution,
as we were unable to check them explicitly. Examples are
the minimal models $\CM_{p,q}$ perturbed by $\phi_{15}$
when $p$ and $q$ are related as
$4p-q=1$.  For these cases 
the twist $\alpha'$ is equal to $2$, and
a trivial shift of 
$f_{L/R}(\theta)$ 
leaves a system with 
$\alpha'=0$. This prevents us from
tuning in a straightforward way onto the required
ultraviolet model: numerically, the massless NLIE
simply recovers 
a flow from $c_{\rm eff}(0)=1$ to 
$c_{\rm eff}(\infty)=1$\,.

We also observe  that (\ref{fl1}) formally
predicts the following flows
from unitary minimal models:
\eq
\CM_{p,p+1} +\phi_{21} \to \CM_{1,p} ~,
\quad(\xi,\alpha')=(\fract{p}{p+1},1)~.
\label{m1p}
\en
However it is a little hard to decide what is meant by $\CM_{1,p}$\,,
and indeed, while
our numerical results
indicate that these
flows behave as expected for small $r$, at some intermediate
scale $c_{\rm eff}(r)$ appears to have a discontinuity.
This could be linked to $\alpha'$ being equal to $1$. In any event,
we suspect there may be a square root singularity 
though a detailed investigation will have to be left for future
work. Note that by the type II equivalence, a similar phenomenon
occurs for the models $\CM_{p,4p-2}$ perturbed by $\phi_{15}$.
In one case a detailed check on these exceptional flows
can be made with relative
ease:
the Ising model $\CM_{3,4}$
perturbed by $\phi_{21}$.  For real coupling $\lambda$,  
an exact  expression for the (massive) 
effective central charge was given in  \cite{KM}.
Using this we can make an exact prediction for 
$c_{\rm eff}$ in the massless case: we send $r\to i \, r$ 
in the perturbative expansion,
swap the logarithmic bulk term 
for that of the massless model (\ref{bulki}) and 
find the coefficient of the $r^2$ term 
 using (\ref{bulkr2}).
The result is:
 \bea
c_{\rm eff}(r) &=& \frac{1}{2} - \frac{3 r^2}{2 \pi^2} 
\biggl[\ln r -\frac{1}{2}-\ln \pi + \gamma_{E} + \frac{\pi}{2 \sqrt 3} \biggr]
\nn\\[2pt]
&&~~~~~+\frac{6}{\pi} \sum_{k=1}^{\infty} \biggl( \sqrt{ (2k-1)^2\pi^2 -r^2}
-(2k-1)\pi+\frac{r^2}{2\, (2k-1)\pi} \biggr)\qquad
\label{cising}
\eea
where 
 $\gamma_E=0.57721556...$ 
is the Euler-Mascheroni constant.
This formula predicts a square root singularity 
at $r{=}\pi$,
preventing a smooth interpolating flow to the far infrared,
but we can at least 
match against the NLIE for values of $r$ out to this point.
As table~\ref{tabising} 
shows, our numerical results agree well with 
the exact formula (\ref{cising}).

\begin{table}
\begin{center}
\begin{tabular}{ l l l }   
\hline 
$r$  & $ ~~c_{\rm eff}^{NLIE}(r) $ & $ ~~c_{\rm eff}^{exact}(r) $  \\
\hline 
 0.0001  &~~0.500000014242146  &~~0.500000014242144  \\ 
 0.6     &~~0.535668551455508  &~~0.535668551455511  \\
 1.1     &~~0.49940046725090    &~~0.49940046725085\\
 1.4     &~~0.4131296597211     &~~0.4131296597212 \\
 1.8     &~~0.18775453205       &~~0.18775453226 \\
\hline
\end{tabular}
\caption{
\protect{\footnotesize 
Comparison of exact and numerical values for 
$c_{\rm eff}$ for the $\phi_{21}$ perturbation of 
$\CM_{3,4}$.
} }
\label{tabising}
\end{center}  
\end{table}

\resection{A remark on $\phi_{13}$ perturbations}
\label{oldstuff}
In this short section we comment on some features of massless
$\phi_{13}$
perturbations which are revealed when similar methods are used.
In \cite{Zd} numerical results were 
only presented for the zero twist ($c_{\rm eff}(0)=
c_{\rm eff}(\infty)=1$) case, but the same equation can describe 
a minimal model $\CM_{p,q}$
perturbed by $\phi_{13}$.
One simply has to set the
kernel parameter $\zeta$ ($p$ in \cite{Zd}) 
to $p/(q{-}p)$, and the twist $\alpha$ in that paper
to $\pi/p$.

Defining an index $J=q-p$~\cite{La}, $\zeta=p/J$ and the flow from 
$\zeta$ to $\zeta-1$ is
\eq
\CM_{p,p+J} +\phi_{13} \to \CM_{p-J,p} \qquad 
\hbox {attracted via $\phi_{31}$~.}
\label{phi13flow}
\en
Again, the number of sequences for each $J$ 
is given by the 
Euler $\varphi$-function, $\varphi(J)$\,.  
As in the $a_2^{(2)}$ case, we cannot access 
all possible minimal models via this equation.  
The kernel $\phi(\theta)$
in the massless equation of \cite{Zd} has a pole at $ i\pi (\zeta-1)$
which crosses the real $\theta$ axis as $\zeta$ falls below $1$.
As before, this could probably
be overcome using analytic continuation but 
for now we choose $\zeta >1$, requiring $2p>q$, to prevent such 
problems occuring.  We also find a set of models, 
$\CM_{p,2p-1}+\phi_{13}$,
where $\alpha'=1$ and the would-be IR model
is $\CM_{1,p}$.  As in the $a_2^{(2)}$ case, 
$c_{\rm eff}$  suffers some sort of discontinuity
at an intermediate scale
for these flows.

\begin{figure}[ht]
\centerline{
\resizebox{.6\linewidth}{!}%
{\includegraphics{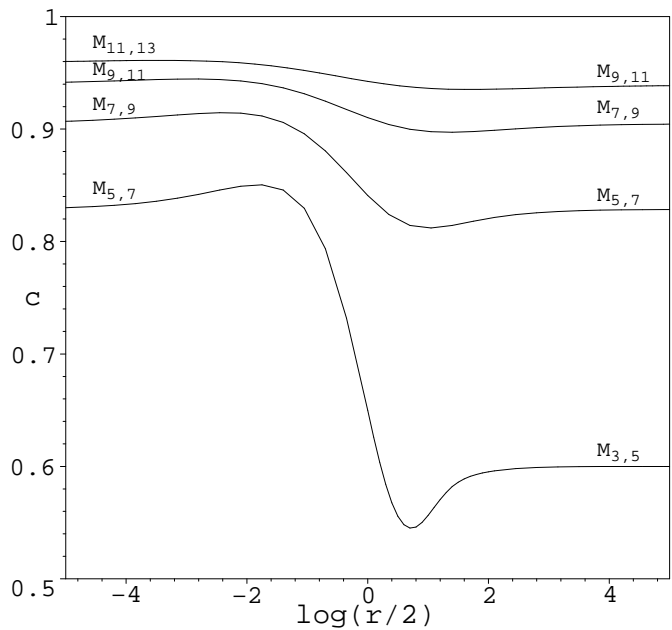}}
}
\vskip 5pt
\caption { \protect { \footnotesize
One of the $J=2$ 
nonunitary $\phi_{13}$ perturbed flows calculated using
the NLIE of \cite{Zd}.
 }}
\label{figjeq2}
\end{figure}

\begin{figure}[ht]
\centerline{
\resizebox{.6\linewidth}{!}%
{\includegraphics{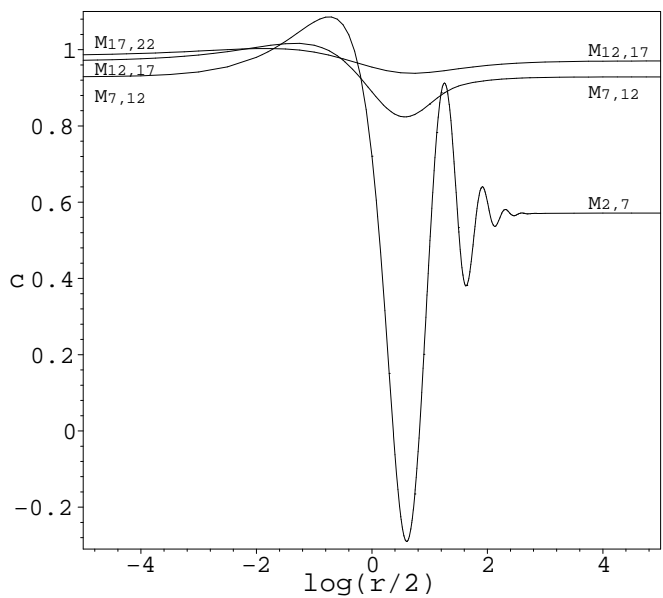}}
}
\vskip 5pt
\caption{\protect {\footnotesize 
One of the nonunitary $\phi_{13}$ perturbed flows with index $J=5$
calculated using
the NLIE of \cite{Zd}.
 }}
\label{figjeq5}
\end{figure}

Solving the massless equation in \cite{Zd} 
for the unitary cases ($J=1$),
we find that results from
the TBA equations of \cite{Zc} are matched
to high accuracy\footnote{we understand 
that this has already been checked by Alyosha Zamolodchikov~\cite{Zpc}}.
The results for $J>1$ 
are perhaps more interesting, since TBA systems for these nonunitary
flows are not known.
Just as for the $\phi_{21}/\phi_{15}$ flows with
$|I|  > 1$, the monotonicity property is lost. Typical cases are
illustrated in figures~\ref{figjeq2}
and \ref{figjeq5}.   


\resection{Conclusions}
\label{conc}
We hope to have demonstrated the rich
structure of flows that can be studied by means of a relatively
simple nonlinear integral equation. A number of unexpected features
have emerged, most notably the consistent failure of the effective
central charges to be monotonic functions of the system size. Indeed,
looking at the full set of minimal 
models, both unitary and nonunitary, we see
that for massless flows a monotonic behaviour of $c_{\rm eff}(r)$ is
very much the 
exception
rather than the rule.

As for future work, the following directions seem natural:
\begin{list}{$\bullet$}{\leftmargin=5pt}
\item
A more detailed study of both the massive and the massless
$a^{(2)}_2$-related nonlinear integral equations is warranted. This
should include a more detailed look at their analytic properties, and
further comparisons with perturbed conformal field theories. In addition,
the equations should be generalised to describe excited states.
\item
The type II conjecture remains a curious observed fact, and a deep
understanding seems still to be lacking.
In cases suffering from this  ambiguity, a knowledge of $c_{\rm
eff}(r)$ will never suffice to disentangle the possible destinations
of the flows. It would be reassuring to confirm that these models
behave as conjectured by some other method.
\item
The work of \cite{Zd} was based in part on
a massless S-matrix. Here we avoided 
this aspect, but it would be interesting to find an S-matrix
description of the new flows. One might also hope to find integrable
lattice models which would yield the massless flows in their continuum
limits.
\item
As discussed in \S3, we were unable to treat 
some flows,
due to singularities in the kernel $\phi(\theta)$
crossing the integration contour.  It would be 
worthwhile to study the NLIE in these zones where $\xi$ falls below 
$1$, perhaps by 
analytic continuation.  It would also be interesting to 
study the NLIE of \cite{Zd}
for $\zeta<1$, where
a similar difficulty is encountered.
\item
The method described in \S2 should enable a massless
nonlinear integral equation to be obtained from any of the (massive)
ADE-related systems described in \cite{Mar,Zi}. It would be interesting to
know whether these share the same strange properties that we have
observed here in the examples related to
$a^{(2)}_2$ and $a^{(1)}_1$.
\item
As mentioned in \S4, the initial discovery of the flows
(\ref{m21flow}) and (\ref{m15flow}) in \cite{Ma} came via a staircase
model. The new flows also fit naturally into staircase-like patterns,
and it is natural to ask whether the existing set of known staircase
models \cite{ZAMSC,Ma,DR1,Ma2,DRg} 
could be enlarged so as to include these cases.
\item
Finally, it remains an open question to find TBA equations
describing the non-unitary $\phi_{13}$ flows, or any of the new
$\phi_{21}$ or $\phi_{15}$ flows.
Why is there no TBA for any non-monotonic case? And what has this to do
with the monstron\footnote{see \cite{Zd}}, if anything?
\end{list}
\medskip

\noindent
{\bf Acknowledgements --} 
We are grateful to Changrim Ahn,  Philippe Di Francesco, 
Francesco Ravanini,
Tom Wynter,
Alyosha Zamolodchikov and Jean-Bernard Zuber
for useful discussions.
PED and TCD thank the EPSRC for an Advanced Fellowship
and a Research Studentship respectively, and RT thanks
the Universiteit van Amsterdam for a post-doctoral
fellowship.
PED and TCD are grateful to SPhT Saclay, and RT to
Durham University and the APCTP, for hospitality during 
various stages of this project.
The work was supported in part by a TMR grant of the
European Commission, reference ERBFMRXCT960012.
%
%
%


\begin{thebibliography}{99}
\raggedright
\parskip 1pt
%
\bibitem{Za}
A.B.\ Zamolodchikov, 
{`Renormalization group and perturbation theory about fixed points in
two-dimensional field theory'},
Sov. J. Nucl. Phys. 46 (1987) 1090\toline{1096}
%
\bibitem{Zb}
A.B.\ Zamolodchikov, 
{`Higher-order integrals of motion in two-dimensional models of the
field theory with a broken conformal symmetry'},
JETP Lett.~46 (1987) 160\toline{164}
%
\bibitem{LC}
A.W.W.\ Ludwig and J.L.\ Cardy,
{`Perturbative evaluation of the conformal anomaly at new critical
points with application to random systems'},
Nucl. Phys. B285 (1987) 687\toline{718}
%
\bibitem{Zc}
Al.B.\ Zamolodchikov, 
{`From tricritical Ising to critical Ising by Thermodynamic Bethe
Ansatz'},
Nucl. Phys. B358 (1991) 524\toline{546}
%
\bibitem{La}
M.\ L\"assig,
{`New hierarchies of multicriticality in two-dimensional field
theory'},
Phys. Lett. B278 (1992) 439\toline{442}
%
\bibitem{Aa}
C.\ Ahn,
{`RG flows of non-unitary minimal CFTs'},
Phys. Lett. B294 (1992) 204\toline{208}
%
\bibitem{Zab}
Al.B.\ Zamolodchikov, 
{`TBA equations for integrable perturbed $SU(2)_k\times SU(2)_l/
SU(2)_{k+l}$ coset models'},
Nucl. Phys. B366 (1991) 122\toline{132}
%
\bibitem{FZa}
V.A.\ Fateev and Al.B.\ Zamolodchikov, 
{`Integrable perturbations of $\ZN$ parafermion models and the $O(3)$
sigma model'},
Phys. Lett. B271 (1991) 91\toline{100}
%
\bibitem{KMspect}
T.R.\ Klassen and E.\ Melzer,
{`Spectral flow between conformal field theories in 1+1 dimensions'},
Nucl. Phys. B370 (1992) 511\toline{550}
%
\bibitem{Maa}
M.J.\ Martins,
{`The thermodynamic Bethe ansatz for deformed $WA_{N-1}$ conformal
field theories'},
Phys. Lett. B277 (1992) 301\toline{305}
%
\bibitem{Ra1}
F.\ Ravanini,
{`Thermodynamic Bethe ansatz for $G_k \times G_l/G_{k+l}$ coset models
perturbed by their $\phi_{1,1,adj}$ operator'},
Phys. Lett. B282 (1992) 73\toline{79} 
%
\bibitem{Ma}
M.J.\ Martins,
{`Renormalization group trajectories from resonance factorized
S matrices'},
Phys. Rev. Lett. 69 (1992) 2461\toline{2464},
{\tt hep-th/9205024};\\
{}~~---~~
{`Exact resonance A-D-E S-matrices and their renormalization group
trajectories'},
Nucl. Phys. B394 (1993) 339\toline{355}
%
\bibitem{RTV1}
F.\ Ravanini, R.\ Tateo and A.\ Valleriani,
{`Dynkin TBAs'},
Int. J. Mod. Phys. A8 (1993) 1707\toline{1727}
%
\bibitem{FOZ}
V.A.\ Fateev, E.\ Onofri  and Al.B.\ Zamolodchikov, 
{`Integrable deformations of the $O(3)$ sigma model. The sausage model'},
Nucl. Phys. B406 (1993) 521\toline{565}
%
\bibitem{FI}
P.\ Fendley and K.\ Intriligator,
{`Exact N=2 Landau-Ginzburg flows'},
Nucl. Phys. B413 (1994) 653\toline{674}
%
\bibitem{FQR}
G.\ Feverati, E.\ Quattrini and F.\ Ravanini,
{`Infrared behaviour of massless integrable flows entering the minimal
models from $\phi_{31}$'},
Phys. lett. B374 (1996) 64\toline{70},
{\tt hep-th/9512104}
%
\bibitem{DTT}
P.\ Dorey, R.\ Tateo and K.E.\ Thompson,
{`Massive and massless phases in self-dual
$\ZN$ spin models: some exact results from
the thermodynamic Bethe ansatz'},
Nucl. Phys. B470 (1996) 317\toline{368},
{\tt hep-th/9601123}
%
\bibitem{RT1}
R.\ Tateo,
{`The sine-Gordon model as $SO(N)_1 \times SO(N)_1/ SO(N)_2$ perturbed
coset theory and generalisation'},
Int. J. Mod. Phys. A10 (1995) 1357\toline{1376}
%
\bibitem{Rac}
F.\ Ravanini, M.\ Stanishkov and R.\ Tateo,
{`Integrable perturbations of CFT with complex parameter:  The $\CM_{3/5}$
model and its generalizations'},
Int. J. Mod. Phys. A11 (1996) 677\toline{698}
%
\bibitem{Zd}
Al.B.\ Zamolodchikov, 
{`Thermodynamics of imaginary coupled 
sine-Gordon. Dense polymer finite-size scaling function'},
Phys. Lett. B335 (1994) 436\toline{443}
%
\bibitem{DDV}
C.\ Destri and H.J.\ de Vega,
{`New thermodynamic Bethe ansatz equations without strings'},
Phys. Rev. Lett. 69 (1992) 2313\toline{2317};\\
{}~~---~~
{`Unified approach to thermodynamic Bethe ansatz and finite size
corrections for lattice models and field theories'},
Nucl. Phys. B438 (1995) 413\toline{454}
%
\bibitem{KBP}
A.\ Kl\"umper, M.T.\ Batchelor and P.A.\ Pearce,
{`Central charges of the 6- and 19-vertex models with twisted
boundary conditions'},
J. Phys. A24 (1991) 3111\toline{3133}
%
\bibitem{FMQR}
D.\ Fioravanti, A.\ Mariottini, E.\ Quattrini and F.\ Ravanini,
{`Excited state Destri-De Vega equation for sine-Gordon
and restricted sine-Gordon models'},
Phys. Lett. B390 (1997) 243\toline{251}, {\tt hep-th/9608091}
%
\bibitem{FRT3}
G.\ Feverati, F.\ Ravanini and  G.\ Takacs,
{`Nonlinear integral equation and finite volume
spectrum of minimal models perturbed by $\phi_{1,3}$'},
{\tt hep-th/9909031}
%
\bibitem{DT1}
P.\ Dorey and R.\ Tateo,
{`Differential equations and integrable models: the $SU(3)$ case'},
{\tt hep-th/9910102}, Nucl. Phys. B, to appear
%
\bibitem{WBN}
S.O.\ Warnaar, M.T.\ Batchelor and B.\ Nienhuis,
{`Critical properties of the Izergin-Korepin and solvable $O(n)$
models and their related quantum spin chains',}
J. Phys. A25 (1992) 3077\toline{3095}
%
\bibitem{SmsG}
F.A.\ Smirnov,
{`Reductions of the sine-Gordon model as a perturbation of 
minimal models of conformal field theory'},
Nucl. Phys. B337 (1990) 156 \toline {180}
%
\bibitem{lecl}
A.\ LeClair,
{`Restricted sine-Gordon theory and the minimal conformal series'},
Phys. Lett. B230 (1989) 103\toline{107}
%
\bibitem{ZamPIII}
Al.B.\ Zamolodchikov, 
{`Painleve III and 2-D Polymers'},
Nucl. Phys. B432 (1994) 427\toline{456}, {\tt hep-th/9409108}
%
\bibitem{Sm}
F.A.\ Smirnov, 
{`Exact S-matrices for $\phi_{12}$-Perturbated minimal models of conformal
field theory'},
Int. J. Mod. Phys. A6 (1991) 1407\toline{1428}
%
\bibitem{MJM}
M.J.\ Martins, 
{`Constructing an S matrix from the truncated conformal
approach  data'},
Phys. Lett. B262 (1991) 39\toline{44} 
%
\bibitem{GT}
G.\ Takacs,
{`A new RSOS restriction of the Zhiber-Mikhailov-Shabat model
and $\Phi_{(1,5)}$ perturbations of nonunitary minimal models'},
Nucl. Phys. B489 (1997) 532\toline{556}, {\tt hep-th/9604098}
%
\bibitem{FSZb}
P.\ Fendley, H.\ Saleur and Al.B.\ Zamolodchikov,
{`Massless flows II: the exact S-matrix approach'},
Int. J. Mod. Phys. A8 (1993) 5751\toline{5778},
{\tt hep-th/9304051}
%
\bibitem{DRg}
P.\ Dorey and F.\ Ravanini,
\ttl{Generalising the staircase models}
Nucl. Phys. B406 (1993) 708\toline{726}
%
\bibitem{Zamy}
Al.B.\ Zamolodchikov,
{`On the thermodynamic Bethe ansatz equations for the reflectionless ADE
scattering theories'},
Phys. Lett. B253 (1991) 391\toline{394}
%
\bibitem{MelzerSUSY}
E.\ Melzer,
{`Supersymmetric analogs of the Gordon-Andrews identities, and related TBA 
systems'},
hep-th/9412154
%
\bibitem{KTW}
H.\ Kausch, G.\ Takacs and G.\ Watts, 
{`On the relation between $\phi_{(1,2)}$ and $\phi_{(1,5)}$ perturbed
minimal models'},
Nucl. Phys. B489 (1997) 557\toline{579}, 
{\tt hep-th/9605104}
%
\bibitem{BCDS}
H.W.\ Braden, E.\ Corrigan, P.E.\ Dorey and R.\ Sasaki, 
{`Affine Toda field
theory and exact S-matrices'}, 
Nucl. Phys. B338 (1990) 689\toline{746}
%
\bibitem{D}
P.\ Dorey, 
{`Root systems and purely elastic S-matrices'},
Nucl. Phys. B358 (1991) 654\toline{676}
%
\bibitem{KMS}
D.A. \ Kastor, E.J. \ Martinec and S.H. \ Shenker,
{`RG flow in N=1 Discrete series'},
Nucl. Phys. B316 (1989) 590\toline{608}
%
\bibitem{FI1}
P.\ Fendley and K.\ Intriligator,
{`Scattering and thermodynamics in integrable N=2 theories'},
Nucl. Phys. B380 (1992) 265
%
\bibitem{FSZa}
P.\ Fendley, H.\ Saleur and Al.B.\ Zamolodchikov,
{`Massless flows I: the sine-Gordon and O(N) models'},
Int. J. Mod. Phys. A8 (1993) 5717\toline{5750},
{\tt hep-th/9304050}
%
\bibitem{ZRSOS}
Al.B.\ Zamolodchikov,
{`Thermodynamic Bethe Ansatz for RSOS scattering theories'},
Nucl. Phys. B358 (1991) 497\toline{523}
%
\bibitem{ZTBA}
Al.B.\ Zamolodchikov,
{`Thermodynamic Bethe ansatz in relativistic models: scaling 3-state Potts
and Lee-Yang models'},
Nucl. Phys. B342 (1990) 695\toline{720}
%
\bibitem{KM}
T.R.\ Klassen and E.\ Melzer,
{`The thermodynamics of purely elastic scattering theories and 
conformal perturbation theory'},
Nucl. Phys. B350 (1991) 635-689
%
\bibitem{FT}
V.A.\ Fateev,
{`The exact relations between the coupling constant 
and the masses of particles for the integrable perturbed conformal
field theories'},
Phys. Lett. B234 (1994) 45\toline{51}
%
\bibitem{FLZZ}
V.\ Fateev, S.\ Lukyanov, Al.B.\ Zamolodchikov and A.B.\ Zamolodchikov,
{`Expectation values of local fields in Bullough-Dodd model 
and integrable perturbed conformal field theories'},
Nucl. Phys. B516 (1998) 652\toline{674}
%
\bibitem{HF}
D.A.\ Huse and M.E.\ Fisher,
{`The decoupling point of the axial next-nearest- neighbour Ising model
and marginal crossover'}, J. Phys. C15 (1982) L585\toline{595}
%
\bibitem{CM}
J.L.\ Cardy and G.\ Mussardo, 
{`Universal properties of self-avoiding walks
from two-dimensional field theory'}, 
Nucl. Phys. B410 (1993) 451\toline{493},
{\tt hep-th/9306028}
%
\bibitem{FD}
Vl.S.\ Dotsenko and V.A.\ Fateev,
{`Operator algebra of two-dimensional conformal theories with 
central charge $c \le 1$'},
Phys. Lett. B154 (1985) 291\toline{295}
%
\bibitem{Zpc}
Al.B.\ Zamolodchikov, private communication
\bibitem{Mar}
A.\ Mariottini, 
{`Ansatz di Bethe termodinamico ed equazione di Destri-de Vega
in teorie di campo bidimensionali'},
Degree thesis (in Italian), Bologna March 1996
%
\bibitem{Zi}
P.\ Zinn-Justin,
{`Nonlinear integral equations for complex Affine Toda models
associated to simply laced Lie algebras'},
J. Phys. A31 (1998) 6747 \toline {6770}
%
\bibitem{ZAMSC}
Al.\ B.\ Zamolodchikov,
{`Resonance factorized scattering and roaming trajectories'},
preprint ENS-LPS-335, 1991
%
\bibitem{DR1}
P.\ Dorey and F.\ Ravanini,
{`Staircase models from Affine Toda Field Theory'},
Int. J. Mod. Phys. A8 (1993) 873\toline{894} 
%
\bibitem{Ma2}
M.J.\ Martins,
{`Analysis of asymptotic conditions in resonance functional
hierarchies'},
Phys. Lett. B304 (1993) 111\toline{114}
%
\end{thebibliography}
\end{document}